\documentclass[fleqn,usenatbib]{mnras}

\usepackage[T1]{fontenc}
\usepackage{ae,aecompl}

\usepackage{graphicx}	
\usepackage{amsmath}	
\usepackage[usenames,dvipsnames]{xcolor}
\usepackage{ragged2e}
\usepackage{ulem}
\usepackage{newtxtext,newtxmath}

\newcommand{\hinvMsun}{h^{-1}{\rm M}_\odot}
\newcommand{\Msun}{{\rm M}_\odot}
\newcommand{\Mh}{M_{\rm h}}
\newcommand{\Mtx}{M_{\rm tx}}
\newcommand{\Mty}{M_{\rm ty}}
\newcommand{\hinvMpc}{h^{-1}{\rm Mpc}}
\newcommand{\Ha}{H$\alpha$}

\newcommand{\Ac}{A_{\rm c}}
\newcommand{\Lc}{L_{\rm c}}
\newcommand{\sigc}{\sigma_{\rm c}}
\newcommand{\Lgap}{\Delta\log L_{\rm cs}}
\newcommand{\phis}{\phi_{\rm s}^*}
\newcommand{\alphas}{\alpha_{\rm s}}

\newcommand{\Acp}{A_{\rm c,p}}
\newcommand{\Lcp}{L_{\rm c,p}}
\newcommand{\phisp}{\phi_{\rm s,p}^*}
\newcommand{\gammaA}{\gamma_A}
\newcommand{\gammaL}{\gamma_L}
\newcommand{\gammaphi}{\gamma_\phi}

\newcommand{\numden}{n_{\rm g}}
\newcommand{\bg}{b_{\rm g}}

\title[Galaxy Clustering with 3D-HST]{The \Ha\ luminosity and stellar mass dependent clustering of star-forming galaxies at $0.7 < z < 1.5$ with 3D-HST}

\author[Clontz, Wake, \& Zheng]{
Callie Clontz$^{1}$\thanks{E-mail: callie.clontz@utah.edu},
David Wake$^{2}$\thanks{E-mail: dwake@unca.edu},
and
Zheng Zheng$^{1}$\thanks{E-mail: zhengzheng@astro.utah.edu}
\\
$^{1}$Department of Physics and Astronomy, University of Utah, 201 James Fletcher Building 115 S. 1400 E., Salt Lake City, UT 84112, USA\\
$^{2}$Department of Physics and Astronomy, University of North Carolina Asheville, 1 University Heights, Asheville, NC 28804, USA\\
}

\date{Accepted XXX. Received YYY; in original form ZZZ}

\pubyear{2022}

\begin{document}
\label{firstpage}
\pagerange{\pageref{firstpage}--\pageref{lastpage}}
\maketitle

\begin{abstract}
We present measurements of the dependence of the clustering amplitude of galaxies on their star formation rate (SFR) and stellar mass ($M_*$) at $0.7 < z < 1.5$ to assess the extent to which environment affects these properties. While these relations are well determined in the local universe, they are much more poorly known at earlier times. For this analysis we make use of the near-IR HST WFC3 grism spectroscopic data in the five CANDELS fields obtained as part of the 3D-HST survey. We make projected 2-point correlation function measurements using $\sim$6,000 galaxies with accurate redshifts, $M_*$ and \Ha\ luminosities. We find a strong dependence of clustering amplitude on \Ha\ luminosity and thus SFR. However, at fixed $M_*$, the clustering dependence on \Ha\ luminosity is largely eliminated. We model the clustering of these galaxies within the Halo Occupation Distribution framework using the conditional luminosity function model and the newly developed conditional stellar mass and \Ha\ luminosity distribution model. These show that galaxies with higher SFRs tend to live in higher mass haloes, but this is largely driven by the relationship between SFR and $M_*$. Finally, we show that the small residual correlation between clustering amplitude and \Ha\ luminosity at fixed $M_*$ is likely being driven by a broadening of the SFR-$M_*$ relationship for satellite galaxies. 
\end{abstract}



\section{Introduction}

\begin{figure*}
  \includegraphics[width=1.0\textwidth,height=0.1475\textheight]{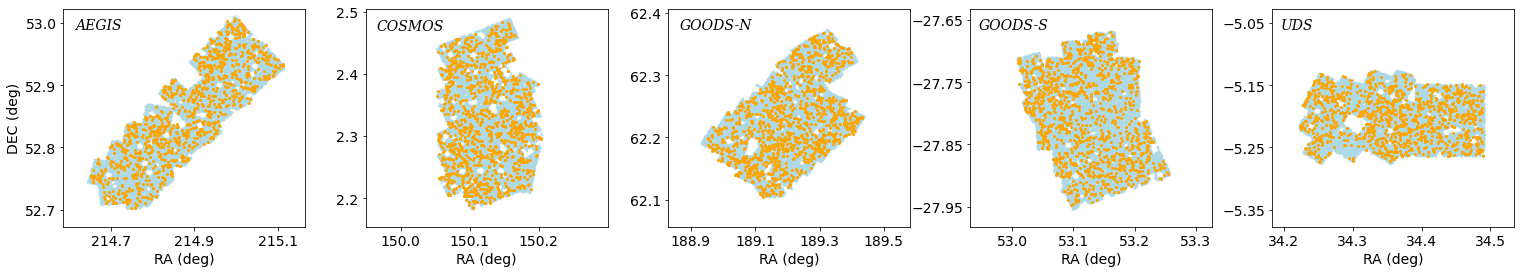}
  \caption{Footprints on the sky of the five survey fields. In each panel, the light blue region shows the footprint, with holes from bright star masks. The orange dots are \Ha-emitting galaxies.
  }
  \label{fig:cosmos_mask}
\end{figure*}

In the cold dark matter (CDM) paradigm of structure formation, the tiny matter density fluctuations resulting from quantum fluctuations during inflation grow over time under the influence of gravitational interactions. Following the formation of dark matter haloes, gas is accreted, cools down, and forms stars to produce galaxies. Galaxies are a natural tracer of the underlying matter distribution to study cosmology. In particular, emission line galaxies, which are mainly star-forming galaxies, have become important targets in current and planned galaxy surveys, such as SDSS-IV eBOSS \citep{Dawson16}, DESI \citep{DESI16}, and surveys with the Roman Space Telescope (RST; \citealt{RST15}) and Euclid \citep{Euclid11}. In this work, we measure the clustering of \Ha-emitting galaxies from Hubble Space Telescope (HST) observations and model the measurements to study the relation between these galaxies and dark matter haloes.

Galaxies are a biased tracers of the underlying matter density field that is connected to cosmology. The galaxy bias encodes information about galaxy formation processes, and a good understanding of it can help tighten cosmological constraints with galaxy clustering. It can be described at the level of individual dark matter haloes as a relation between galaxies and haloes. Such a halo occupation distribution (HOD) formalism specifies how galaxies of a given sample occupy dark matter haloes, which can be constrained by galaxy clustering data (such as the widely used measurements of the two-point correlation function, hereafter the 2PCF).

It has been shown that galaxy properties, including star formation rate (SFR) and stellar mass ($M_*$), are correlated with their environment \citep{blanton} and a deeper look into this relationship will facilitate a better understanding of galaxy formation and evolution. The finer details that describe the extent to which environmental processes affect galaxy evolution are still unknown. Processes such as feedback from star formation, black hole accretion, as well as gas cooling each act on various timescales and have different dependencies on galaxy mass and environment. One way to characterise the environment dependence of galaxy properties is to measure the dependence of galaxy clustering on galaxy properties. The HOD modelling of the clustering measurements connects galaxies to their natural environment, the dark matter haloes. The dependence of galaxy clustering on galaxy properties (such as luminosity and colour) has been intensively studied \citep[e.g.][]{zheng_coil,Zehavi11}. 

The primary goal of this research is to use clustering of star-forming galaxies to investigate the role environment plays in determining the specific SFR (sSFR) and stellar mass of galaxies. It has been shown in the local universe that SFR depends strongly on environment, where galaxies with lower sSFR are found in denser environments \citep{blanton} and that the clustering amplitude has a strong dependence on SFR \citep[e.g.][]{li08}. In this work, we use the data from 3D-HST with well-defined redshifts and photometry to analyse the clustering of \Ha-emitting galaxies and to study the connection between SFR and stellar mass and dark matter haloes at an earlier time.

Our work is similar to previous works by \citet{cochrane17} and \citet{cochrane18}, who studied the clustering of $\sim 4,000$ \Ha\ emitters at $z = 0.8$, 1.47, and 2.23 using the narrow-band High-Redshift(Z) Emission Line Survey (HiZELS). We perform similar measurements with the 3D-HST grism spectroscopic survey at $0.7 < z < 1.5$ with the common goal of tracking the evolution of the clustering dependence on galaxy properties. The measurements and modelling results provide useful inputs to plan future surveys, such as those with RST \citep{RST15} and Euclid \citep{Euclid11}.

The paper is organised as follows. In Section~\ref{sec:data}, we introduce the 3D-HST data and the construction of various \Ha-emitting galaxy samples. In Section~\ref{sec:2PCF_measurement}, we describe the 2PCF measurements. In Section~\ref{sec:2PCF_powerlaw}, the dependence of 2PCFs on \Ha\ luminosity and stellar mass is presented and characterised. The halo-based models are introduced and applied to model the 2PCF measurements in Section~\ref{sec:hod}, which include a conditional luminosity function (CLF) model to interpret the dependence of 2PCFs on \Ha\ luminosity and a conditional stellar mass and \Ha\ luminosity distribution model to interpret the joint dependence of 2PCFs on stellar mass and \Ha\ luminosity. We compare our results with previous work in Section~\ref{sec:discussion} and conclude in Section~\ref{sec:conclusion}.

Throughout this paper, in carrying out the clustering measurements, we adopt a spatially-flat $\Lambda$CDM cosmology with density parameters $\Omega_{\rm m} = 0.3$ and $\Omega_\Lambda = 0.7$. In modelling the clustering, we further adopt the following cosmological parameters, $\Omega_{\rm b}=0.048$, $H_0 = 100h {\rm km\, s^{-1}Mpc^{-1}}$ with $h=0.68$, $n_{\rm s}=0.96$, and $\sigma_8=0.81$. Haloes are defined as bound regions with mean density 200 times that of the background universe.

\section{Data}
\label{sec:data}

\begin{table*}
 \caption{\Ha\ Luminosity-Bin and Luminosity-Threshold Galaxy Samples}
 \begin{tabular}{||c c c c l r r r c c c||} 
 \hline
 Sample & $\log L_{\rm H\alpha,min}$ & $\log L_{\rm H\alpha,max}$  & $\log \langle L_{\rm H\alpha}\rangle $ & $\langle z \rangle $ & $N_{\rm g}$ & $N_{\rm g, weighted}$ & $\numden$ & $r_0$ & $\gamma$ & $r_{0} \ [\gamma_{\rm med}]$\\ 
 \hline
LB1 & 41.10 & 41.30 & 41.21 & 0.82 & 565 & 3942 & 10.29 & $1.34 \pm 0.62$ & $1.17 \pm 0.09$ & $2.03 \pm 0.19$ \\
LB2 & 41.30 & 41.50 & 41.40 & 0.92 & 1162 & 3492 & 9.12 & $2.15 \pm 0.47$ & $1.28 \pm 0.08$ & $2.30 \pm 0.22$ \\
LB3 & 41.50 & 41.70 & 41.60 & 1.04 & 1380 & 2222 & 5.80 & $2.39 \pm 0.45$ & $1.35 \pm 0.09$ & $2.24 \pm 0.21$ \\
LB4 & 41.70 & 41.90 & 41.80 & 1.14 & 1251 & 1520 & 3.97 & $3.24 \pm 0.27$ & $1.55 \pm 0.09$ & $2.52 \pm 0.32$ \\
LB5 & 41.90 & 42.10 & 41.99 & 1.17 & 792 & 932 & 2.43 & $3.66 \pm 0.18$ & $1.62 \pm 0.06$ & $2.64 \pm 0.29$ \\
LB6 & 42.10 & 42.50 & 42.24 & 1.21 & 559 & 646 & 1.69 & $4.43 \pm 0.89$ & $1.26 \pm 0.11$ & $4.70 \pm 0.51$ \\

\hline \hline
LB1Lz & 41.10 & 41.30 & 41.21 & 0.82 & 565 & 1675 & 3.39 & $1.34 \pm 0.62$ & $1.17 \pm 0.09$ & $2.11 \pm 0.04$ \\
LB2Lz & 41.30 & 41.50 & 41.40 & 0.91 & 1094 & 1517 & 3.19 & $1.83 \pm 0.50$ & $1.24 \pm 0.08$ & $2.24 \pm 0.04$ \\
LB3Lz & 41.50 & 41.70 & 41.60 & 0.93 & 834 & 1003 & 2.18 & $2.77 \pm 0.48$ & $1.33 \pm 0.08$ & $2.81 \pm 0.06$ \\
LB4Lz & 41.70 & 41.90 & 41.80 & 0.93 & 560 & 671 & 1.46 & $4.12 \pm 0.45$ & $1.53 \pm 0.13$ & $3.59 \pm 0.23$ \\
LB5Lz & 41.90 & 42.10 & 42.00 & 0.93 & 295 & 349 & 0.77 & $4.37 \pm 0.38$ & $1.53 \pm 0.11$ & $3.78 \pm 0.18$ \\
LB6Lz & 42.10 & 42.50 & 42.23 & 0.95 & 173 & 199 & 0.45 & $5.04 \pm 1.26$ & $1.34 \pm 0.16$ & $4.99 \pm 0.30$ \\

LB1Hz & 41.50 & 41.70 & 41.63 & 1.23 & 546 & 662 & 1.43 & $1.33 \pm 1.04$ & $1.27 \pm 0.26$ & $1.70 \pm 0.04$ \\
LB2Hz & 41.70 & 41.90 & 41.80 & 1.30 & 691 & 825 & 1.80 & $2.33 \pm 0.30$ & $1.40 \pm 0.09$ & $2.30 \pm 0.04$ \\
LB3Hz & 41.90 & 42.10 & 41.99 & 1.31 & 497 & 583 & 1.30 & $2.99 \pm 0.42$ & $1.59 \pm 0.16$ & $2.45 \pm 0.06$ \\
LB4Hz & 42.10 & 42.50 & 42.25 & 1.33 & 386 & 447 & 1.01 & $4.31 \pm 0.48$ & $1.38 \pm 0.10$ & $4.33 \pm 0.23$ \\

\hline \hline
 LT1 & 41.10 & 42.50 & 41.63 & 0.97 & 5709 & 12757 & 33.30 & $2.82 \pm 0.16$ & $1.44 \pm 0.04$ & $2.84 \pm 0.15$ \\
LT2 & 41.30 & 42.50 & 41.73 & 1.04 & 5144 & 8815 & 23.01 & $3.29 \pm 0.15$ & $1.51 \pm 0.03$ & $3.30 \pm 0.15$ \\
LT3 & 41.50 & 42.50 & 41.86 & 1.11 & 3982 & 5322 & 13.89 & $3.22 \pm 0.23$ & $1.48 \pm 0.05$ & $3.22 \pm 0.22$ \\
LT4 & 41.70 & 42.50 & 41.99 & 1.16 & 2602 & 3099 & 8.09 & $3.72 \pm 0.17$ & $1.50 \pm 0.05$ & $3.72 \pm 0.16$ \\
LT5 & 41.90 & 42.50 & 42.11 & 1.19 & 1351 & 1579 & 4.12 & $4.32 \pm 0.36$ & $1.50 \pm 0.09$ & $4.32 \pm 0.34$ \\
LT6 & 42.10 & 42.50 & 42.24 & 1.21 & 559 & 646 & 1.69 & $4.43 \pm 0.89$ & $1.26 \pm 0.11$ & $4.73 \pm 0.55$ \\

 \hline \hline 
LT1Lz & 41.10 & 42.50 & 41.60 & 0.89 & 3521 & 5417 & 11.44 & $2.97 \pm 0.17$ & $1.40 \pm 0.04$ & $2.98 \pm 0.02$ \\
LT2Lz & 41.30 & 42.50 & 41.70 & 0.93 & 2956 & 3741 & 8.05 & $3.51 \pm 0.18$ & $1.46 \pm 0.03$ & $3.46 \pm 0.04$ \\
LT3Lz & 41.50 & 42.50 & 41.83 & 0.93 & 1862 & 2223 & 4.86 & $3.85 \pm 0.29$ & $1.41 \pm 0.05$ & $3.84 \pm 0.07$ \\
LT4Lz & 41.70 & 42.50 & 41.96 & 0.94 & 1028 & 1220 & 2.68 & $4.10 \pm 0.37$ & $1.47 \pm 0.08$ & $3.95 \pm 0.12$ \\
LT5Lz & 41.90 & 42.50 & 42.10 & 0.94 & 468 & 548 & 1.22 & $4.19 \pm 0.47$ & $1.39 \pm 0.08$ & $4.27 \pm 0.11$ \\
LT6Lz & 42.10 & 42.50 & 42.23 & 0.95 & 173 & 199 & 0.45 & $2.08 \pm 4.14$ & $1.12 \pm 0.27$ & $4.56 \pm 0.47$ \\

LB1Hz & 41.50 & 41.70 & 41.63 & 1.23 & 546 & 662 & 1.43 & $1.33 \pm 1.04$ & $1.27 \pm 0.26$ & $1.70 \pm 0.12$ \\
LB2Hz & 41.70 & 41.90 & 41.80 & 1.30 & 691 & 825 & 1.80 & $2.33 \pm 0.30$ & $1.40 \pm 0.09$ & $2.30 \pm 0.04$ \\
LB3Hz & 41.90 & 42.10 & 41.99 & 1.31 & 497 & 583 & 1.30 & $2.99 \pm 0.42$ & $1.59 \pm 0.16$ & $2.45 \pm 0.15$ \\
LB4Hz & 42.10 & 42.50 & 42.25 & 1.33 & 386 & 447 & 1.01 & $4.31 \pm 0.48$ & $1.38 \pm 0.10$ & $4.33 \pm 0.16$ \\

\hline 
 \end{tabular}

From top to bottom are four sets of galaxy samples -- the \Ha-luminosity-bin (LB) samples,  the LB samples at lower and higher redshifts (Lz and Hz), the luminosity-threshold (LT) samples, and the LT samples at lower and higher redshifts (Lz and Hz). 

 For each sample, shown are the minimum, maximum, and mean \Ha\ luminosity (in units of ${\rm erg\, s^{-1}}$), mean redshift, total number of galaxies, the sum of the combined weights, and galaxy number density (in units of $10^{-3}h^3{\rm Mpc}^{-3}$), correlation length ($r_0$) and power-law index ($\gamma$) from the power-law 2PCF fit. The last column shows the value of $r_0$ with $\gamma$ fixed to the median value. 
\label{tab:lum_bin_thres}
\end{table*}

\begin{table*}
 \caption{Galaxy Samples in Bins of Stellar Mass and \Ha\ Luminosity}
 \begin{tabular}{||c r r r c c c c c r c c||} 
 \hline
 Sample & $\log M_{\rm *,min}$ & $\log M_{\rm *,max} $ & $\langle \log M_{*} \rangle$ & $\log L_{\rm H\alpha,min} $ & $\log L_{\rm H\alpha,max} $ & $\log \langle L_{\rm H\alpha}\rangle $ & $\langle z \rangle $ & $N_{\rm g}$ & $N_{\rm g, weighted}$ & $\numden$ & $r_0 [\gamma_{\rm fixed}]$ \\
 \hline
LB1MB1 & 9.20 & 9.44 & 9.31 & 41.10 & 41.64 & 41.39 & 0.90 & 429 & 1296 & 3.38 & $3.81 \pm  0.32$ \\ 
LB1MB2 & 9.44 & 9.78 & 9.59 & 41.10 & 41.64 & 41.45 & 0.93 & 442 & 1035 & 2.70 & $3.99 \pm 0.33$\\ 
LB1MB3 & 9.78 & 11.50 & 10.26 & 41.10 & 41.64 & 41.45 & 0.94 & 429 & 886 & 2.31 & $5.67 \pm 0.58$\\ 
LB2MB1 & 9.20 & 9.55 & 9.38 & 41.64 & 41.90 & 41.75 & 1.17 & 425 & 531 & 1.39 & $3.67 \pm 0.32$\\ 
LB2MB2 & 9.55 & 9.97 & 9.73 & 41.64 & 41.90 & 41.77 & 1.11 & 451 & 549 & 1.43 & $3.25 \pm 0.54$\\ 
LB2MB3 & 9.97 & 11.50 & 10.42 & 41.64 & 41.90 & 41.78 & 1.09 & 437 & 518 & 1.35 & $5.25 \pm 0.45$\\ 
LB3MB1 & 9.20 & 9.84 & 9.58 & 41.90 & 42.50 & 42.08 & 1.20 & 402 & 468 & 1.22 & $4.02 \pm 0.61$\\ 
LB3MB2 & 9.84 & 10.31 & 10.06 & 41.90 & 42.50 & 42.12 & 1.17 & 417 & 486 & 1.27 & $3.56 \pm 0.64$\\ 
LB3MB3 & 10.31 & 11.50 & 10.65 & 41.90 & 42.50 & 42.16 & 1.18 & 403 & 466 & 1.22 & $6.26 \pm 0.46$\\ 

\hline \hline 
MB1LB1 & 9.20 & 9.57 & 9.36 & 41.10 & 41.52 & 41.34 & 0.88 & 420 & 1443 & 3.77 & $2.70 \pm 0.39$\\ 
MB1LB2 & 9.20 & 9.57 & 9.39 & 41.52 & 41.73 & 41.63 & 1.08 & 441 & 658 & 1.72 & $2.08 \pm 0.32$\\ 
MB1LB3 & 9.20 & 9.57 & 9.41 & 41.73 & 42.50 & 41.94 & 1.19 & 425 & 495 & 1.29 & $3.27 \pm 0.28$\\ 
MB2LB1 & 9.57 & 10.06 & 9.77 & 41.10 & 41.67 & 41.48 & 0.95 & 426 & 884 & 2.31 & $3.74 \pm 0.52$\\ 
MB2LB2 & 9.57 & 10.06 & 9.79 & 41.67 & 41.90 & 41.79 & 1.11 & 441 & 528 & 1.38 & $2.55 \pm 0.37$\\ 
MB2LB3 & 9.57 & 10.06 & 9.83 & 41.90 & 42.50 & 42.10 & 1.18 & 424 & 493 & 1.29 & $2.98 \pm 0.68$\\ 
MB3LB1 & 10.06 & 11.50 & 10.49 & 41.10 & 41.77 & 41.52 & 0.97 & 422 & 736 & 1.92 & $5.53 \pm 0.49$\\ 
MB3LB2 & 10.06 & 11.50 & 10.47 & 41.77 & 42.04 & 41.91 & 1.13 & 440 & 515 & 1.34 & $3.49 \pm 0.59$\\ 
MB3LB3 & 10.06 & 11.50 & 10.50 & 42.04 & 42.50 & 42.22 & 1.19 & 378 & 439 & 1.15 & $6.33 \pm 0.66$\\ 

 \hline \hline
 
M1L1 & 9.10 & 9.60 & 9.32 & 40.99 & 41.60 & 41.36 & 0.89 & 771 & 2514 & 6.56 & -\\ 
M1L2 & 9.10 & 9.60 & 9.37 & 41.60 & 41.90 & 41.74 & 1.14 & 682 & 876 & 2.29 & -\\ 
M1L3 & 9.10 & 9.60 & 9.39 & 41.90 & 43.05 & 42.10 & 1.22 & 235 & 274 & 0.72 & -\\ 
M2L1 & 9.60 & 10.10 & 9.80 & 40.99 & 41.60 & 41.42 & 0.93 & 298 & 712 & 1.86 & -\\ 
M2L2 & 9.60 & 10.10 & 9.82 & 41.60 & 41.90 & 41.75 & 1.09 & 547 & 681 & 1.78 & -\\ 
M2L3 & 9.60 & 10.10 & 9.86 & 41.90 & 43.05 & 42.12 & 1.18 & 449 & 524 & 1.37 & -\\ 
M3L1 & 10.10 & 11.50 & 10.50 & 40.99 & 41.60 & 41.38 & 0.92 & 223 & 526 & 1.37 & -\\ 
M3L2 & 10.10 & 11.50 & 10.53 & 41.60 & 41.90 & 41.77 & 1.07 & 373 & 447 & 1.17 & -\\ 
M3L3 & 10.10 & 11.50 & 10.52 & 41.90 & 43.05 & 42.21 & 1.18 & 630 & 732 & 1.91 & -\\ 

\hline 
 \end{tabular}

From top to bottom are three sets of galaxy samples. The first set of samples are constructed by cuts in \Ha\ luminosity bin (LB), and within each LB sample, galaxies are further divided into stellar mass bin (MB) samples. These samples are fitted with $\rm \gamma = 1.45$. The second set of samples are constructed by cuts in stellar mass (MB), and within each MB sample, galaxies are further divided into \Ha\ luminosity bin  (LB) samples. These samples are fitted with $\rm \gamma = 1.31$. The third set of samples are similar to the second set, but with a central \Ha\ luminosity-bin sample and \Ha\ luminosity-threshold samples for the upper and lower luminosity samples. No power laws are fitted to these samples. The 2PCF measurements from the third set are used to constrain the conditional stellar mass and \Ha\ luminosity distribution model. 
 For each sample, shown are the minimum and maximum $M_*$ (in units of $\Msun$), and mean $\log M_*$, minimum, maximum, and mean \Ha\ luminosity (in units of ${\rm erg\, s^{-1}}$), mean redshift, total number of galaxies, the sum of the combined weights, and galaxy number density (in units of $10^{-3}h^3{\rm Mpc}^{-3}$).
\label{tab:mass_bin_lum_sample}
\end{table*}

In this work, we make use of near-IR HST WFC3 grism spectroscopic data in the five CANDELS fields obtained as part of the 3D-HST survey (\citealp[]{brammer}; \citealp[]{momcheva}; \citealp[]{skelton}). 
This survey, carried out by the Hubble Space Telescope in 2015, consists of deep near-infrared low resolution WFC3/G141 grism spectroscopic measurements, which are combined with existing deep multi-wavelength photometry to yield accurate redshifts ($\sigma_z/(1+z) \sim 0.0003$), stellar masses, and emission line measurements. The G141 grism provides a spectral resolution $R\sim$130 and covers the wavelength range of $1.1\ {\rm \mu m}$ to $1.65\ {\rm \mu m}$.   
The 3D-HST footprints of the five fields (AEGIS, COSMOS, GOODS-N, GOODS-S, and UDS) cover a total area of $\sim$625 arcmin$^2$, and are shown in Fig.~\ref{fig:cosmos_mask} together with the \Ha-emitting galaxies used in our analysis.

In this work we are interested in the \Ha\ emission line which is well detected in the G141 grism spectra over a redshift range of $0.7 < z < 1.5$, setting a natural redshift range for our analysis. The $5\sigma$ limiting flux of the 3D-HST G141 observations is $\sim 3\times 10^{-17} {\rm erg\,s^{-1}\,cm^{-2}}$ \citep[]{brammer}. 

 To define our master sample of \Ha\ emitting galaxies we select all galaxies with $0.7 < z < 1.5$, with clean photometry (use\_phot=1), a grism redshift, jh\_mag < 26, and \Ha\ flux $>4\times 10^{-17} {\rm erg\,s^{-1}\,cm^{-2}}$. This redshift range corresponds to the wavelength range of the G141 grism where \Ha\ falls and the sensitivity remains high. The other cuts ensure a clean sample of galaxies with robust \Ha\ detections. 

To obtain the \Ha\ luminosity of each galaxy, we make a correction to remove the contamination from the adjacent [N\,{\small II}] emission lines that are not separately resolved in the grism spectra. We use the stellar mass -- [N\,{\small II}]/\Ha\ flux ratio (stellar mass--metallicity relation) in \citet{wuyts},
\begin{equation}
    \log\left({\rm [N\,{\scriptstyle II}]}/{\rm H\alpha}\right)
    = \frac{1}{0.57}\left\{ -0.1+\log\left[1-\exp\left(-\left[\frac{M_*}{10^{10.2}\Msun}\right]^{0.4}\right) \right] \right\}.
\end{equation}
In our galaxy sample, the median of the correction factor $1/(1+{\rm [N\, {\scriptstyle II}]}/{\rm H\alpha})$ is about $-0.06$ dex.

While our sample of \Ha\ emitting galaxies is highly complete in order to accurately measure the clustering we need to account for two observational effects. Firstly, the \Ha\ flux limit means that our \Ha\ luminosity limit increases with increasing redshift. We could deal with this by defining volume-limited samples however that removes a significant fraction of the sample that is already fairly small for 2PCF measurements. Instead we apply a $V_{\rm max}$ weight to be used in pair counting for effectively volume-limited measurements.  

For both the pair counts in the clustering measurements and number density measurements, we assign to each galaxy a 1/$\Delta V_{\rm max}$ weight, which is the comoving volume within which a galaxy is detectable, given its intrinsic luminosity, our sample flux limit and the redshift cuts applied to a given sub-sample. For each galaxy, we use its \Ha\ luminosity and our sample flux limit to obtain the maximum redshift $z_{\rm max}$ at which that galaxy would still be included in the sample. For a sample with a redshift range $z_1<z<z_2$, we define $z_{\rm lower}=z_1$ and $z_{\rm upper}={\rm min}\{z_2,\, z_{\rm max}\}$ for a given galaxy and compute the comoving volume $\Delta V_{\rm max}$ between $z_{\rm lower}$ and $z_{\rm upper}$. The $V_{\rm max}$ weight is then defined as $1/\Delta V_{\rm max}$.

The second effect concerns an increasing level of incompleteness for galaxies close on the sky as a result of overlapping grism spectra. Such contamination is modelled and corrected for in the 3D-HST pipeline, however severely contaminated spectra result in failed grism redshift measurements \citep{brammer}. Such a scale dependent incompleteness can lead to an underestimate in the correlation function on scales where this is important. To correct for this effect we reassign the weight of any galaxy that failed to get a grism redshift evenly between all the galaxies with grism redshifts that lie within 100 $h^{-1}$kpc.

\begin{figure}
  \includegraphics[width=0.5\textwidth]{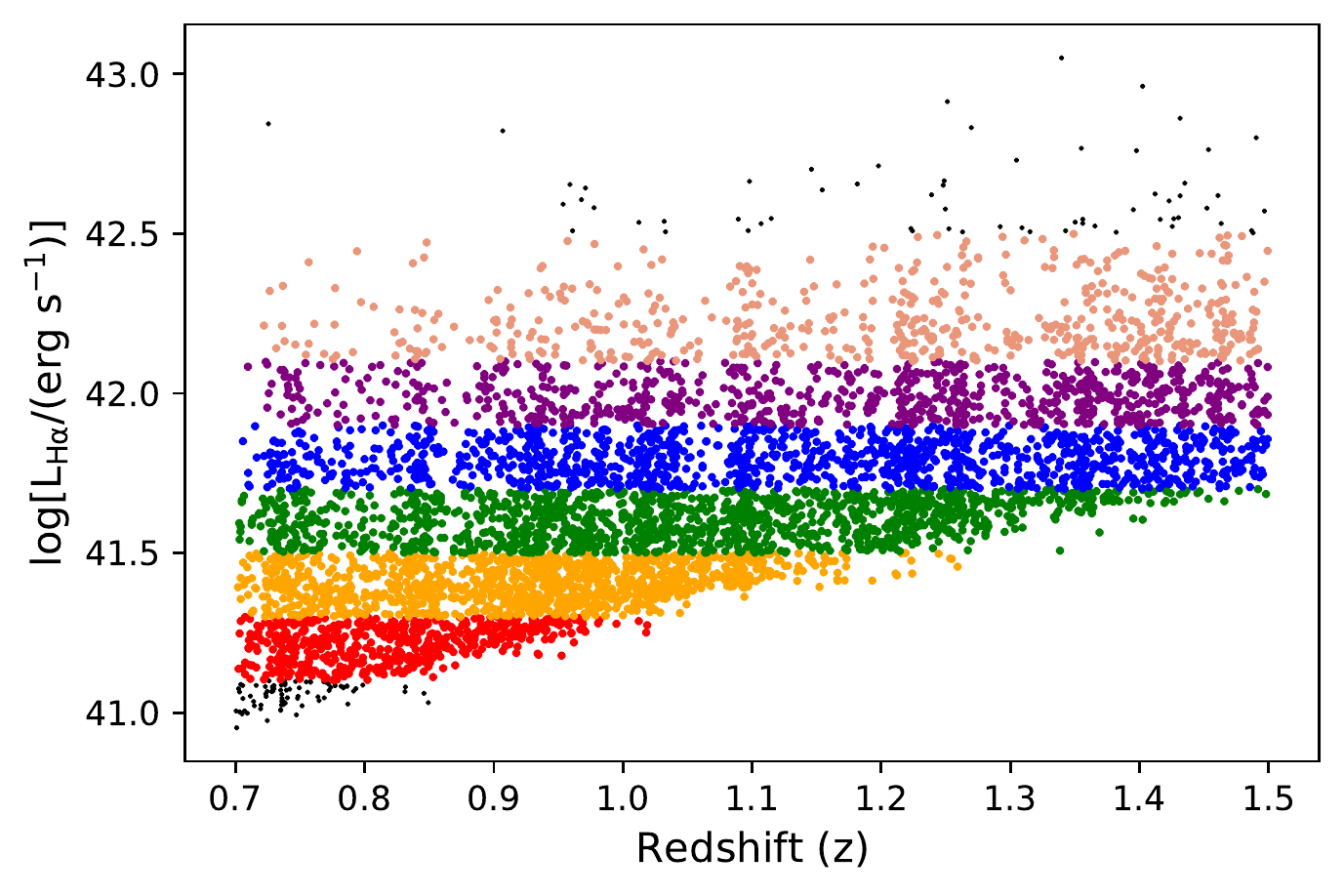}
  \caption{Luminosity-bin samples and redshift ranges. Galaxies in different \Ha\ luminosity bins are coded with different colours. We choose samples with equal width in \Ha\ luminosity except the uppermost sample which extends to $\log L_{\rm H\alpha}=42.5$. Effectively volume-limited 2PCF measurements are achieved from these flux-limited samples through proper $V_{\rm max}$ weights.
  }
  \label{fig:lum_bin_Lha_vs.z}
\end{figure}

In the redshift range of $0.7<z<1.5$, we construct both luminosity-bin and luminosity-threshold samples of galaxies. The details of the samples can be found in Table~\ref{tab:lum_bin_thres}.
Fig.\ref{fig:lum_bin_Lha_vs.z} shows the luminosity and redshift ranges of the luminosity-bin samples.
To study the evolution of clustering and compare to previous work, we also construct luminosity-bin and luminosity-threshold samples in two additional redshift range, $0.7<z<1.1$ and $1.1<z<1.5$.
We also construct galaxy samples in fine bins of \Ha\ luminosity and stellar mass to study the joint dependence of clustering on both quantities, which will be presented in Section~\ref{sec:stellar_mass}.

Finally, for measuring the clustering of galaxies, we create a random catalogue for each sample. The angular distribution of the random points matches the footprints of the five survey fields, taking into account the offset in the dispersion of the grism in the spectral direction, and the masking of regions around bright stars (Fig.~\ref{fig:cosmos_mask}). For the radial distribution of the random points, we make use of the galaxy sample. For each random point, we randomly draw a galaxy from the sample, obtaining its $z_{\rm lower}$ and $z_{\rm upper}$. The random point follows a uniform distribution within the comoving volume between $z_{\rm lower}$ and $z_{\rm upper}$, from which its radial comoving distance (hence its redshift) is drawn. The random point is assigned the $1/\Delta V_{\rm max}$ weight and a unity collision weight.

For each galaxy sample, the number of random points is typically 150 times that of the galaxies.

\section{Two-Point Correlation Function Measurements}
\label{sec:2PCF_measurement}

\begin{figure*}
  \includegraphics[width=1.0\textwidth]{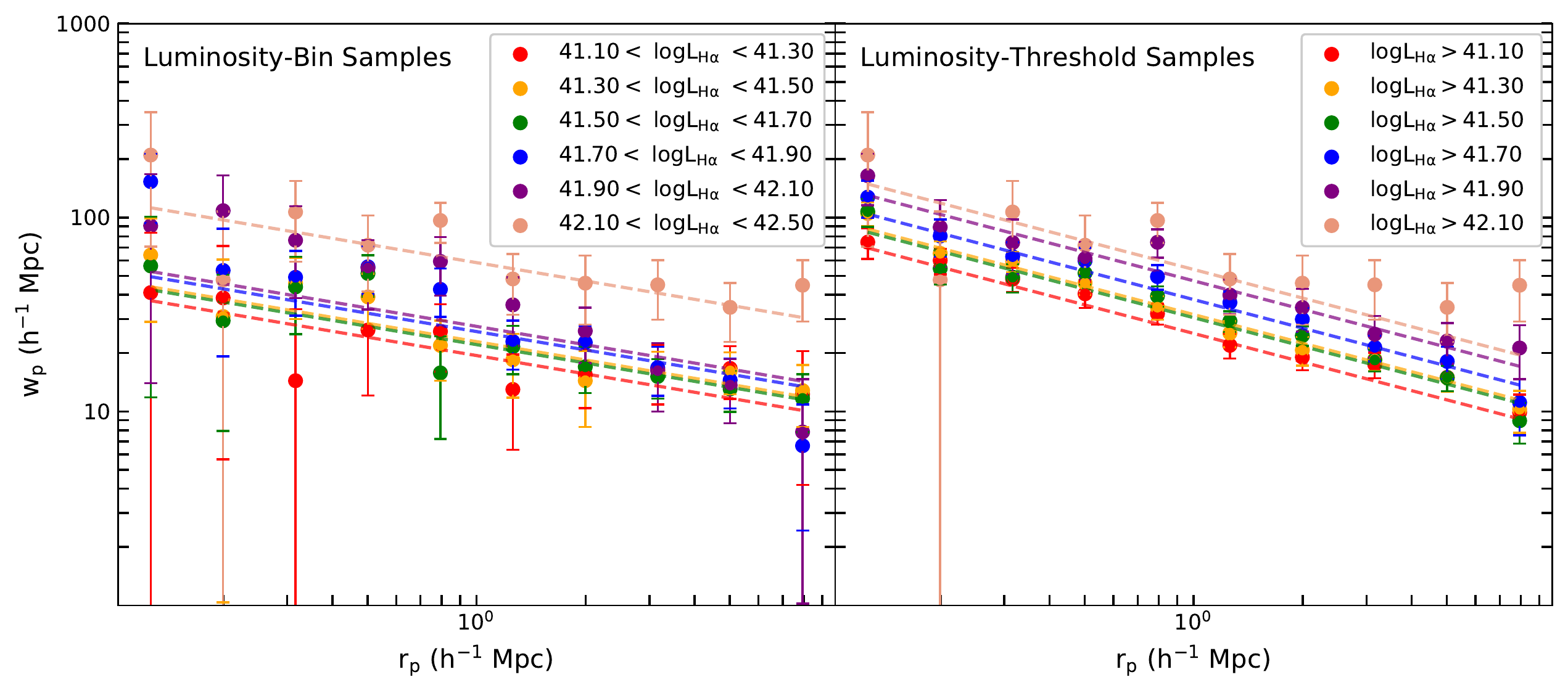}
  \caption{Projected 2PCFs and power-law fits for luminosity-bin samples (left) and luminosity-threshold samples (right). $L_{\rm H\alpha}$ is given in units of $\rm erg \ s^{-1}$.}
  \label{fig:pwr_law_lum_bins}
\end{figure*}

The 2PCF is a statistical measure of the excess probability of finding, in this case, two galaxies at a given separation when compared to a uniformly distributed random sample. We quantify the clustering of each sample of 3D-HST galaxies using the projected 2PCF. 

First, we measure the redshift-space 2PCF for each sample using the Landy-Szalay estimator \citep{LS93}, as a function of transverse pair separation $r_p$ and line-of-sight pair separation $r_\pi$,
\begin{equation}
\xi(r_p, r_\pi) = \frac{{\rm DD} - 2{\rm DR} + {\rm RR}}{\rm RR},
\label{eqn:LS}
\end{equation}
where DD, DR, and RR are the data-data, data-random, and random-random pair counts within a given $(r_p, r_\pi)$ pair separation bin, normalized by the corresponding total numbers of pairs, respectively. We set logarithmic bins for $r_p$ centered at $\log [r_p/(\hinvMpc)]=-1$ to 1 with bin width $\Delta \log r_p = 0.20$, and linear bins for $r_\pi$ from 0 to 50$\hinvMpc$ with bin width $\Delta r_\pi$=0.05 $\hinvMpc$.

The redshift-space 2PCF $\xi(r_p, r_\pi)$ is projected along the line-of-sight direction to obtain the projected 2PCF $w_p$, which reduces the redshift-space distortion effect \citep{Davis83}. We have
\begin{equation}
    w_p(r_p) = 2\int_0^{r_{\pi,{\rm max}}} \xi(r_p, r_\pi) =  2\sum_i \xi(r_p, r_{\pi,i})\Delta r_\pi,
\end{equation}
where $r_{\pi,i}$ is the $i$-th $r_\pi$ bin and we take $r_{\pi,{\rm max}}=50\hinvMpc$.

Specifically, we adopt the code Super W of Theta (SWOT; \citealt{coupon}) for our 2PCF calculations. As we construct each \Ha\ luminosity-bin sample from the flux-limited survey, the maximum redshift depends on luminosity. Each galaxy or random point is assigned the $1/\Delta V_{\rm max}$ weight and collision weight. Following \citet{Xu16}, we modify the SWOT code so that during the pair counting, each pair is given the larger $1/\Delta V_{\rm max}$ weight of the two objects, in addition to the product of the collision weights. That is, the DD, DR, and RR pair counts in equation~(\ref{eqn:LS}) are all with such weights.

The 1/$\Delta V_{\rm max}$ weight ensures that the clustering measurements are effectively volume-limited \citep[e.g.][]{Xu16}. For each sample of galaxies, the covariance matrix of the projected 2PCF measurement is estimated from 128 jackknife samples.

If the real-space 2PCF $\xi(r)$ is approximated by a power law, 
\begin{equation}
    \xi(r)=\left(\frac{r}{r_0}\right)^{-\gamma},
    \label{eqn:xi_pl}
\end{equation}
the clustering strength $r_0$ and the power-law index $\gamma$ can be obtained from the power-law fit to $w_p$,
\begin{equation}
  w_{p}(r_p) = r_p \left(\frac{r_p}{r_0} \right)^{-\gamma} \ \frac{\Gamma(\gamma/2 -1/2) \ \Gamma(1/2)}{\Gamma(\gamma/2)}.
  \label{eqn:wp_pl}
\end{equation}
We will first present the clustering results characterised by the power-law fits, before performing the physically motivated halo modelling. 

Finally, given the limited volume of the 3D-HST survey, the integral constraint (IC), resulting from taking the measured galaxy number density to be the global mean, may become appreciable on large scales. We estimate the IC in the real-space 2PCF for each galaxy sample by making use of the random catalogue and a model real-space 2PCF,
\begin{equation}
    C_\xi = \frac{\sum_i  \xi^{\rm mod}(r_i) {\rm RR}(r_i)}{\sum_i {\rm RR}(r_i)},
\end{equation}
which leads to the corresponding IC in the projected 2PCF
\begin{equation}
    C = 2 \int_0^{r_{\pi, {\rm max}}} C_\xi \,{\rm d}r_\pi = 2C_\xi r_{\pi, {\rm max}}.
\end{equation}
In the expression, ${\rm RR}(r_i)$ is the number of random-random pairs with separation in the $r_i$ bin (linearly spaced bins).
The largest separation bin in our calculation is essentially set by the line-of-sight range ($\sim 3000\hinvMpc$) of the sample volume. 

For the model 2PCF $\xi^{\rm mod}(r)$, we use that derived from the power-law fit to $w_p$. We apply a cutoff in $\xi^{\rm mod}(r)$ beyond 60$\hinvMpc$ to mimic the trend seen in the matter 2PCF, but we verify that this cutoff has little effect in the derived IC. The IC value is then added to $w_p$ to refine the power-law fit and the IC estimate. The procedure is iterated to reach a converged IC. We find that for each sample the value of IC is of the order of $1\hinvMpc$. This is about 20--35 per cent of the error bars of the data points at the largest scale ($\sim 8\hinvMpc$) in our $w_p$ measurements, not a significant effect. 

\section{Dependence of Galaxy Clustering on \Ha\ Luminosity and Stellar Mass}
\label{sec:2PCF_powerlaw}

\begin{figure*}
  \includegraphics[width=1.0\textwidth]{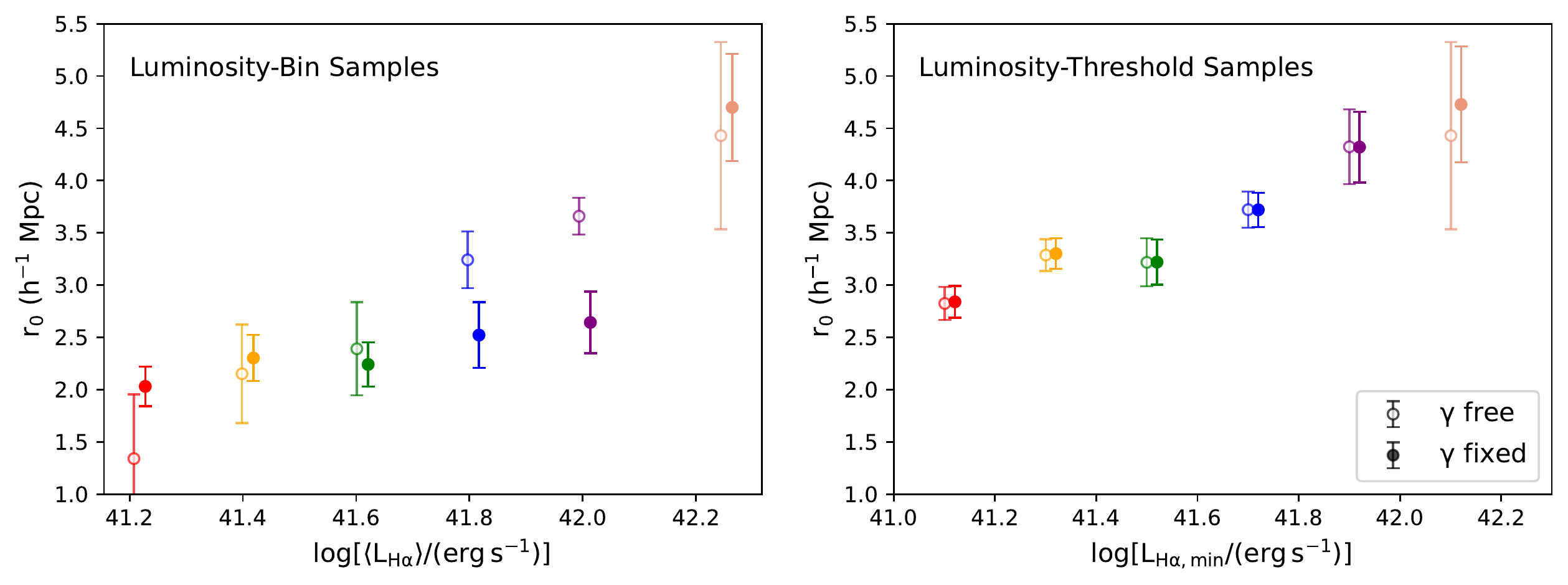}
  \caption{Luminosity dependent clustering strength ($r_0$) for luminosity-bin (left) and luminosity-threshold (right) samples. In each panel, open circles are from power-law fits by keeping the index $\gamma$ as a free parameter, and filled circles are derived by fixing $\gamma$ to the median value of all the samples (see text).
}
  \label{fig:r0_lum}
\end{figure*}

In this Section, we present the clustering measurements of \Ha-emitting galaxies. Following the common practice, we characterise the dependence of clustering strength on \Ha\ luminosity and stellar mass based on power-law fits to the projected 2PCFs. In the next section, we will interpret the clustering in a more physical and informative way using the halo model.

\subsection{Dependence of Clustering on \Ha\ Luminosity}

The left and right panels of Fig.~\ref{fig:pwr_law_lum_bins} show the projected 2PCF measurements of our luminosity-bin and luminosity-threhold samples, respectively. Overall, galaxies with higher \Ha\ luminosity are more strongly clustered. For the luminosity-bin samples, there is also evidence for the more \Ha\ luminous galaxies having steeper clustering profiles on small scales (below $\sim 0.3\hinvMpc$). 

To characterise the clustering amplitude we fit the $w_p$ measurement of each sample with a power law and obtain the correlation length $r_0$ and the power-law index $\gamma$ (equations~\ref{eqn:xi_pl} and \ref{eqn:wp_pl}). To better compare the clustering amplitude among different samples, we also redo the power-law fits with index $\gamma$ fixed to the median value of the $\gamma$ free fits, which is 1.31 and 1.49 for the luminosity-bin and luminosity-threshold samples, respectively. These fits are shown as dashed lines in Fig.~\ref{fig:pwr_law_lum_bins}. While a single power law appears to underestimate the clustering amplitude on small scales ($<0.2\hinvMpc$) for the more luminous luminosity-bin samples, it provides a reasonable description of the large-scale clustering for all the samples.

In Fig.~\ref{fig:r0_lum} we plot $r_0$ as a function of \Ha\ luminosity for the luminosity-bin (left) and luminosity-threshold (right) samples with both $\gamma$ free and fixed cases. In both figures there is a clear dependence of $r_0$, hence the clustering amplitude, on \Ha\ luminosity, with the lowest luminosity samples having the weakest clustering.

\subsection{Dependence of Clustering on \Ha\ Luminosity and Stellar Mass}
\label{sec:stellar_mass}

\begin{figure}
  \includegraphics[width=0.5\textwidth]{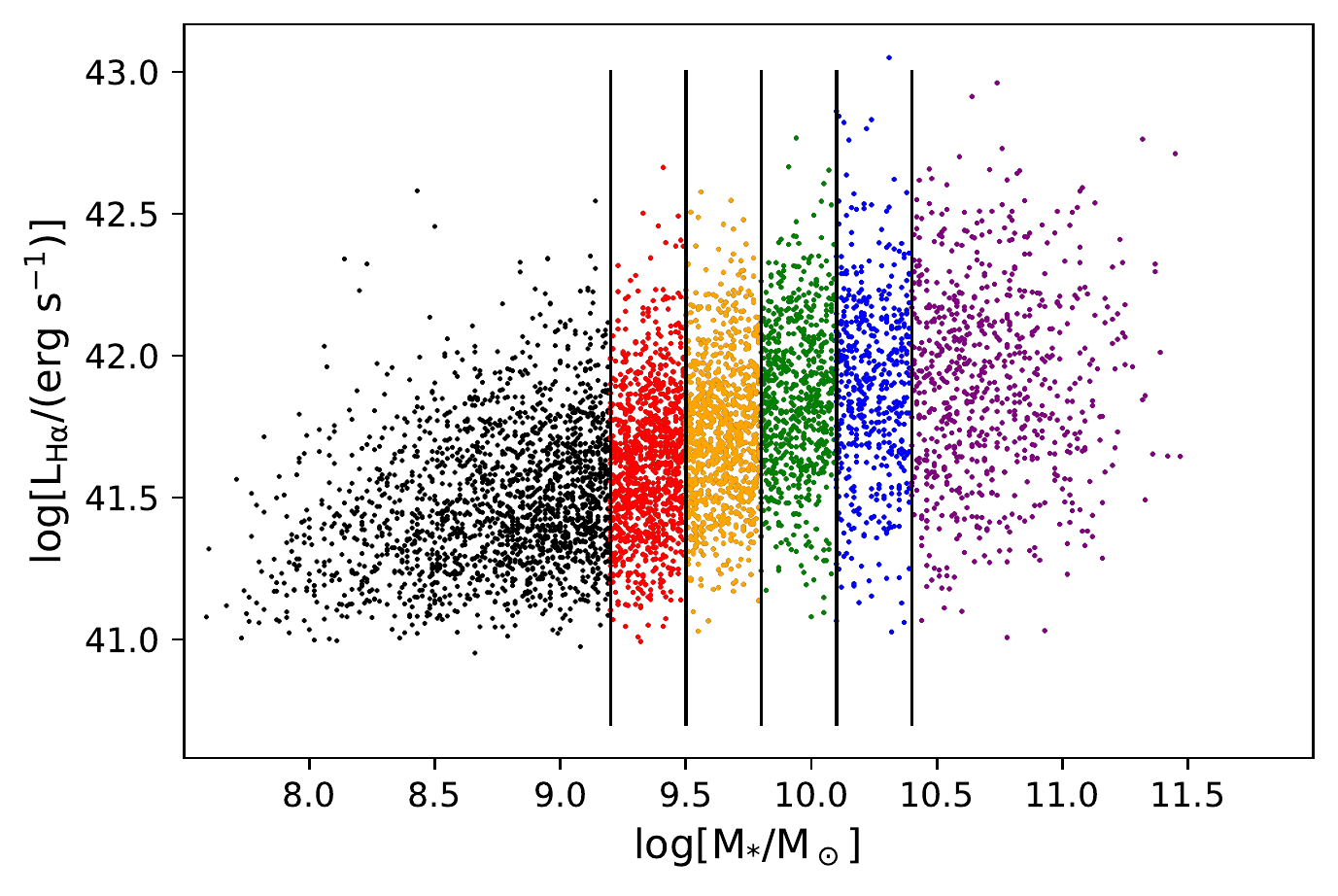}
  \caption{Construction of galaxy samples in five stellar mass bins. The vertical lines delineate the range of the stellar mass bins.
  }
  \label{fig:stellar_mass_bins}
\end{figure}

\begin{figure*}
  \includegraphics[width=1.0\textwidth]{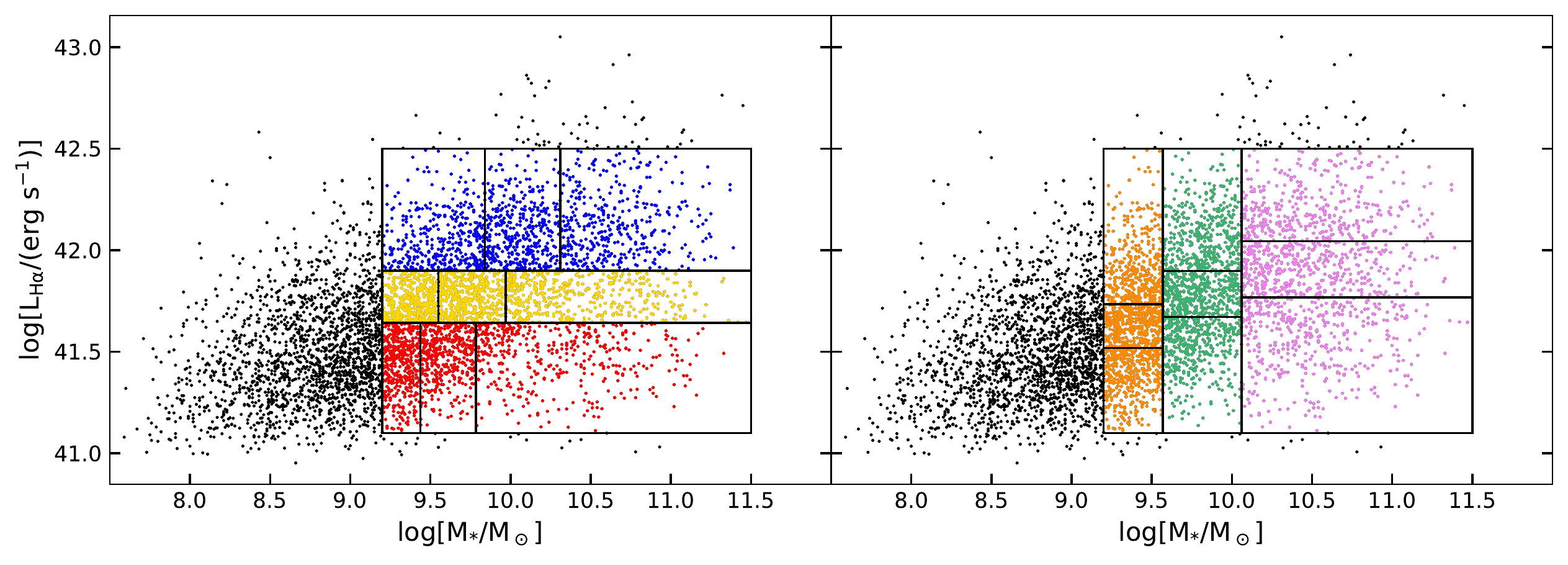}
  \caption{Left: Construction of stellar mass-bin samples at fixed \Ha\ luminosity. The upper limits of each stellar mass bin are chosen to be the 33rd, 66th, and 100th quantiles of galaxies with $\log(M_*/\Msun) > 9.2$, ensuring that each bin has an approximately equal number of galaxies to have good signal-to-noise ratios across our measurements. Right: Construction of \Ha\ luminosity-bin samples at fixed stellar mass using similar quantile binning.}
  \label{fig:bins_fixed_sm_fixed_halum}
\end{figure*}

\begin{figure*}
  \includegraphics[width=1.0\textwidth]{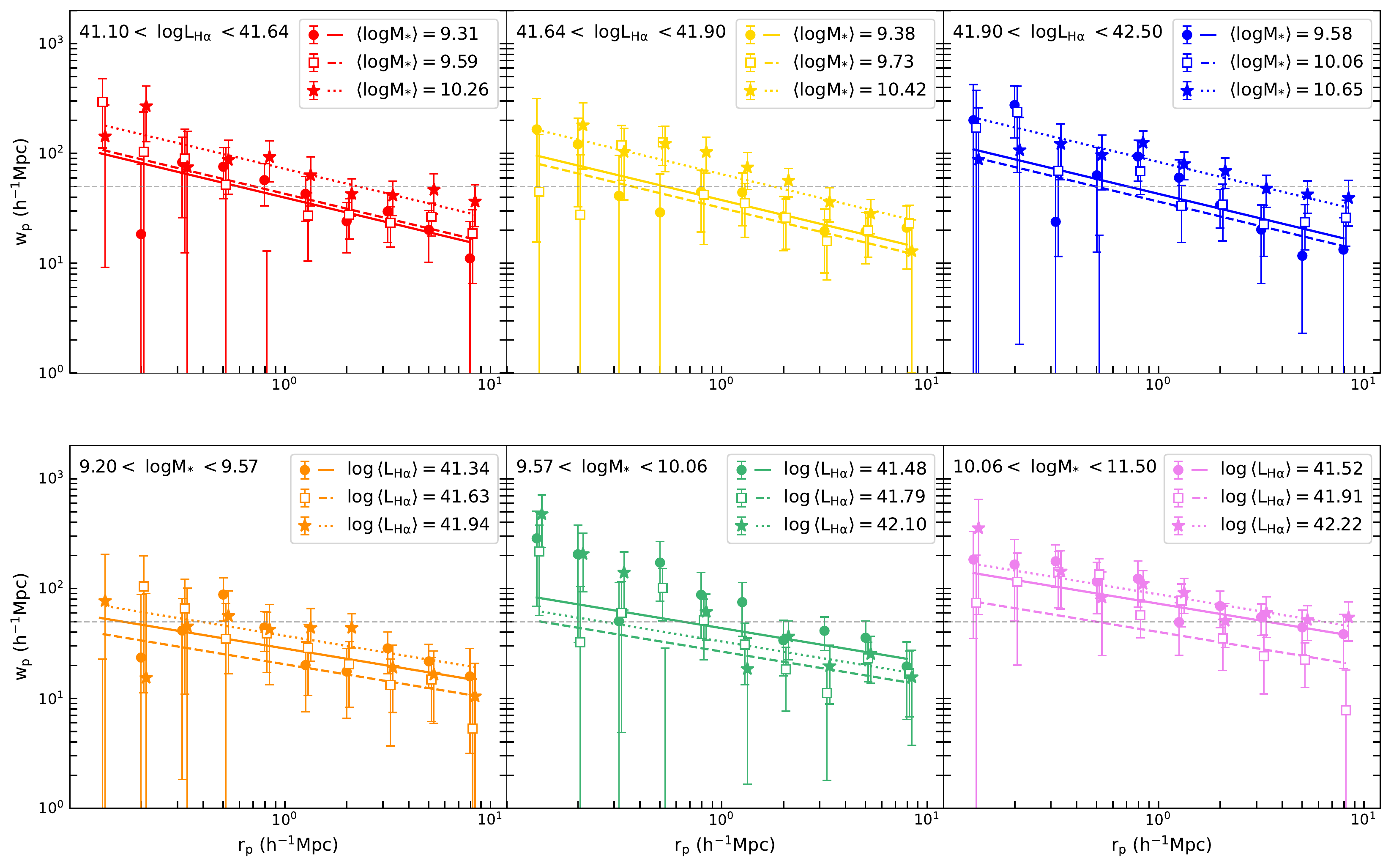}
  \caption{Top row: Projected 2PCFs for stellar mass-dependent samples in bins of \Ha\ luminosity increasing left to right along with power-law models. The slope of the power law fits are fixed to the median value for the $H_{\alpha}$ Luminosity binned samples. Similarly the models across the bottom panels have a fixed slope of median from the stellar mass binned samples. The positions of the points on the $r_{\rm p}$ axis are slightly offset for clarity. In each bin of \Ha\ luminosity the highest stellar mass galaxies show the highest clustering amplitude. There is little clear variation in the clustering amplitude as the \Ha\ luminosity increases from the left to right panel. 
  Bottom row: Same as the top row but now for \Ha\ luminosity-dependent samples in bins of stellar mass. The slope of the power law fits are fixed to the median value for the stellar mass binned samples. There is little variation and no clear trend in the clustering amplitude with \Ha\ luminosity in a given bin in stellar mass, but there is an increase in the clustering amplitude as the stellar mass increases left to right. The dashed light grey horizontal line in each panel is shown to make it easier to compare the clustering amplitude between panels. 
 }
  \label{fig:wp_fixed_sm}
\end{figure*}

\begin{figure*}
  \includegraphics[width=1.0\textwidth]{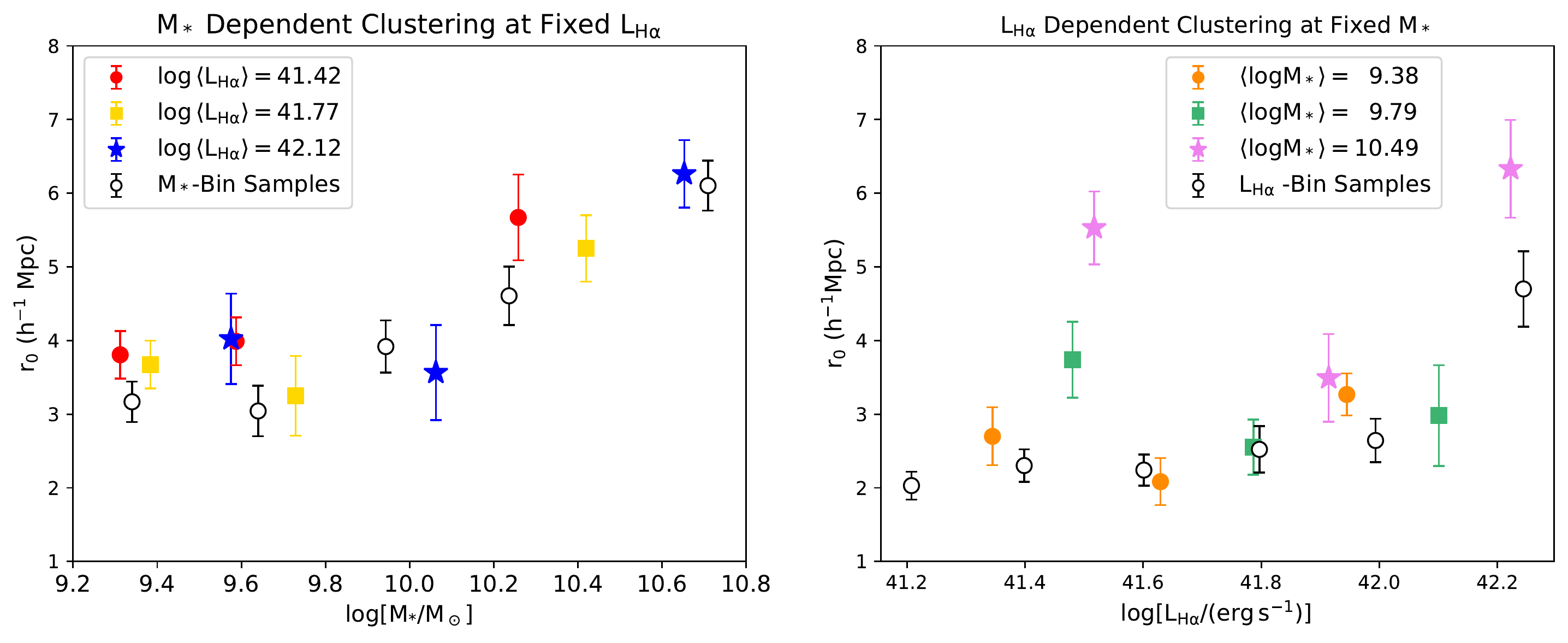}
  \caption{
    Dependence of clustering strength ($r_0$) on stellar mass and \Ha\ luminosity.
    Left: $r_0$ as a function of stellar mass in bins of \Ha\ luminosity. The three colours represent three fixed \Ha\ luminosity bins with each bin split into the highest, lowest and middle thirds by stellar mass. Irrespective of the \Ha\ luminosity the clustering amplitude $r_0$ follows the overall trend with stellar mass, indicating little variation of the clustering with \Ha\ luminosity at fixed stellar mass.
    Right: $r_0$ as a function of \Ha\ luminosity in bins of stellar mass. The three colours represent three fixed stellar mass bins with each bin split into the highest, lowest and middle thirds by \Ha\ luminosity. Again there is at most a weak trend in the \Ha\ luminosity dependent clustering within a stellar mass bin but now there is a clear offset between the stellar mass bins indicating a stronger dependence of clustering amplitude on stellar mass at fixed \Ha\ luminosity. For comparison, the open circles show the dependence of $r_0$ on stellar mass (left) and \Ha\ luminosity (right) inferred from stellar mass-bin and \Ha\ luminosity-bin samples. All samples on the left are fitted with $\rm \gamma = 1.45$. Similarly, all samples on the right are fitted with $\rm \gamma = 1.31$.
    }
  \label{fig:r0_stellar_mass}
\end{figure*}

It has long been known that galaxy clustering depends strongly on galaxy stellar mass \citep[e.g.][]{wake}, and there is a correlation between SFR and stellar mass (i.e. the star formation main sequence; e.g. \citealt{li08}). Here we find that galaxy clustering depends on \Ha\ luminosity (a proxy for SFR). Is this dependence driven by the dependence of clustering on stellar mass and the underlying correlation between SFR and $M_*$? To answer this question, we further divide the galaxies into different stellar mass bins and construct luminosity-dependent galaxy samples within the stellar mass bin.

The construction of the galaxy samples in bins of stellar mass is illustrated in Fig.~\ref{fig:stellar_mass_bins}. Five stellar mass bins (indicated by the five vertical lines) are formed within the range of $9.2< \log(M_*/\Msun) < 11.5$, with an approximately equal number of galaxies in each bin. The lower bound, $\log(M_*/\Msun)=9.2$, is chosen to ensure the completeness in stellar mass, given the survey limit. 

Next, to study the clustering dependence on both stellar mass and \Ha\ luminosity, we used a quantile binning scheme to create samples of equal size at both fixed \Ha\ luminosity and fixed $M_*$ shown in the left and right panels of Fig.~\ref{fig:bins_fixed_sm_fixed_halum}, respectively. See Table~\ref{tab:mass_bin_lum_sample}. 

We follow the same procedure as in Sections~\ref{sec:data} and \ref{sec:2PCF_measurement} to construct random catalogues and measure the projected 2PCF for each sample. Again, we follow the above procedure and perform power-law fits first with the index $\gamma$ as a free parameter and then set to the median value. 

The resulting $w_p$ measurements for the samples in bins of \Ha\ luminosity and $M_*$ are shown in Fig.~\ref{fig:wp_fixed_sm}. For the $M_*$-dependent clustering, we only show the inferred $r_0$ from the power-law fits in Fig.~\ref{fig:r0_stellar_mass} (open symbols in the left panel), which increases with stellar mass as expected.

Looking at the bottom row of Fig.~\ref{fig:wp_fixed_sm} there is no monotonic trend in the clustering amplitude with \Ha\ luminosity when stellar mass is fixed. In each case the middle \Ha\ luminosity bin has the lowest overall clustering amplitude. Conversely, comparing panels left to right shows a general trend of increasing clustering amplitude with increasing stellar mass. Turning to the top row in Fig.~\ref{fig:wp_fixed_sm}, where each panel shows the clustering dependence on stellar mass in a bin of \Ha\ luminosity, the highest stellar mass sample shows a much higher clustering amplitude in each panel. Looking left to right there is a small increase in the clustering amplitude in the highest \Ha\ luminosity bin over the lower two, which show little difference from each other. 

These trends are confirmed by Fig.~\ref{fig:r0_stellar_mass}. The left panel shows the clustering strength ($r_0$) as a function of stellar mass for samples of differing \Ha\ luminosity. There is a clear trend of stronger clustering for galaxies of higher stellar mass above $M_* \sim 10^{10} \Msun$, with a fairly flat trend at lower masses. There is little variation between the three set of samples binned by \Ha\ luminosity with them  all lying close to the same overall trend, with any variation consistent with the uncertainties. This implies that when stellar mass is fixed there is little dependence of the large-scale clustering amplitude on \Ha\ luminosity. In the right panel of Fig.~\ref{fig:r0_stellar_mass} we now show $r_0$ as a function of \Ha\ luminosity for three bins in stellar mass. This time we see something quite different, with no universal trend apparent. The three stellar mass bins are offset from one another and within a given mass bin there is no monotonic trend with \Ha\ luminosity. This demonstrates that stellar mass is the dominant factor in determining the clustering strength of galaxies and implies that the \Ha-luminosity-dependent galaxy clustering is largely a manifestation of the correlation between SFR and stellar mass (a.k.a. the star formation main sequence).

It is worth noting that there is a weak V-shaped trend of $r_0$ with \Ha\ luminosity in each of the stellar mass bin samples, with the middle \Ha\ luminosity bin showing a lower clustering amplitude than the upper and lower luminosity bin. If this trend is real, it suggests that galaxies with both the lowest and highest specific SFRs (sSFR) are more strongly clustered than those in the middle. Such a trend could be in part driven by star-forming satellite galaxies having either reduced or enhanced sSFRs compared to central star-forming galaxies of the same stellar mass. That is, the star formation main sequence of satellite galaxies may have a larger scatter in SFR than that of central galaxies. We provide more insights into this trend in Section~\ref{sec:HOD_lum_mass} based on the halo modelling results.

While a power-law characterisation of the 2PCFs allow us to study the overall dependence of galaxy clustering on \Ha\ luminosity and stellar mass, a halo-based model will provide us a more informative way to interpret the clustering measurements and to study the galaxy-halo connection. We turn to such a model in the following section.

\section[HOD]{Halo Modelling}
\label{sec:hod}

In this section, we model the clustering measurements within the framework of halo occupation distribution (HOD). With an assumed cosmology, the properties of dark matter haloes, including their mass function and spatial clustering, are readily known. The HOD specifies the relationship between galaxies and dark matter haloes as a function of halo mass. In particular, it parameterises the probability distribution function of finding galaxies in haloes of a given mass, including the mean occupation function. Together with the halo population, the HOD model enables us to calculate the 2PCFs for a sample of galaxies to compare to observational measurements.

\subsection{Model Setups}

To model the luminosity-dependent clustering, we parameterise the galaxy-halo relation in terms of the \Ha\ luminosity distribution of galaxies as a function of halo mass. The halo occupation function of each sample can then be derived by applying the sample luminosity cuts. With such a parameterisation, we are able to simultaneously model the clustering measurements of all the samples with different luminosity cuts.

Similarly, to model the luminosity and stellar mass dependent clustering, the galaxy-halo relation is parameterised by the joint distribution of \Ha\ luminosity and stellar mass as a function of halo mass, which allows simultaneous modelling of all the samples of various luminosity and stellar mass cuts.

In this subsection, we present our models for the above two cases, and in the next two subsections we present the modelling results.

\subsubsection{Model for the Luminosity Dependent Clustering }
\label{sec:CLF}

As we study luminosity-dependent clustering with luminosity-bin galaxy samples, it is convenient to parameterise the galaxy-halo relation in terms of the conditional luminosity function (CLF; \citealt{Yang03}), which describes the luminosity distribution of galaxies as a function of halo mass. The mean occupation function related to the HOD for a given galaxy sample can then be obtained by integrating the CLF over the luminosity range used to construct the sample.

We follow the CLF parameterisation in \citet{Yang08} and separate it into contributions from central and satellite galaxies. In haloes of a given mass $\Mh$, the central galaxy CLF is described by a log-normal distribution,
\begin{equation}
  \Phi_{\rm cen}(L|\Mh) \equiv \frac{{\rm d}\langle N_{\rm cen}\rangle}{{\rm d}\log L} = \frac{A_{\rm c}}{\sqrt{2\pi}\sigma_{\rm c}} \exp\left[-\frac{(\log L - \log L_{\rm c})^2}{2\sigma_{\rm c}^2}\right] ,
\end{equation}
where $A_{\rm c}$, $\sigma_{\rm c}$, and $\log L_{\rm c}$ denote the amplitude, width, and the centre of the log-normal function, respectively, with all possibly depending on halo mass. The amplitude $A_{\rm c}$, which is the integral of the above expression over all luminosities, represents the fraction of central galaxies that are star-forming (with \Ha\ emission). By definition, $A_{\rm c}$ cannot exceed unity.

The CLF of satellite galaxies in haloes of a given mass $\Mh$ is parameterised as a Schechter-like function,
\begin{equation}
  \Phi_{\rm sat}(L|\Mh)\equiv \frac{{\rm d}\langle N_{\rm sat}\rangle}{{\rm d}L} = \phi^*_{\rm s}\left(\frac{L}{L^*_{\rm s}}\right)^{\alpha_{\rm s}} \exp\left[-\left(\frac{L}{L^*_{\rm s}}\right)^2\right].
\end{equation}
The form deviates from the Schechter function by having a squared term inside the exponential function, which better describes the empirically determined CLF from a galaxy group catalogue \citep{Yang08}. Here $\phi_{\rm s}^*$, $\alpha_{\rm s}$, and $L_{\rm s}^*$ are the normalisation, power-law slope at the low-luminosity end, and the characteristic cutoff luminosity. In our model, we parameterise $L_{\rm s}^*$ through the luminosity gap between central and satellite galaxies, defined as $\Delta \log L_{\rm cs} = \log L_{\rm c} - \log L_{\rm s}^*$.

With the CLF, for a sample of galaxies with luminosity in the range $L_1<L<L_2$, the mean occupation functions of central and satellite galaxies in haloes of mass $\Mh$ can be computed as
\begin{eqnarray}
  \langle N_{\rm cen}(\Mh)\rangle  = \int_{L_1}^{L_2} \Phi_{\rm cen}(L|\Mh) d\log L \qquad\qquad\qquad\qquad  \label{eqn:Ncen}\\
  =\frac{A_{\rm c}}{2}\left[{\rm erf}\left(\frac{\log L_2-\log L_{\rm c}}{\sqrt{2}\sigma_{\rm c}}\right) - {\rm erf}\left(\frac{\log L_1-\log L_{\rm c}}{\sqrt{2}\sigma_{\rm c}}\right)\right],
\end{eqnarray}
and
\begin{eqnarray}
  \langle N_{\rm sat}(\Mh)\rangle  =  \int_{L_1}^{L_2} \Phi_{\rm sat}(L|\Mh) d L \qquad\qquad\qquad\qquad\qquad \label{eqn:Nsat}\\
   =  \frac{\phi_{\rm s}^*L_{\rm s}^*}{2}\left[\gamma\left(\frac{\alpha_{\rm s}+1}{2},\left(\frac{L_2}{L_{\rm s}^*}\right)^2\right)- \gamma\left(\frac{\alpha_{\rm s}+1}{2},\left(\frac{L_1}{L_{\rm s}^*}\right)^2\right) \right], \label{eqn:Nsat2}
\end{eqnarray}
where ${\rm erf}(x)=(2/\sqrt{\pi})\int_0^x \exp(-t^2){\rm d}t$ is the error function and $\gamma(s, x)=\int_0^x t^{s-1}\exp(-t){\rm d}t$ is the lower incomplete gamma function. The expression in equation~(\ref{eqn:Nsat2}) is for the case with $\alpha_{\rm s}>-1$. For $\alpha_{\rm s}\leq -1$, the integral in equation~(\ref{eqn:Nsat}) is evaluated numerically.

In haloes of a given mass, the above CLF forms are described by parameters $\Ac$, $\Lc$, $\sigc$, $\phis$, $\alphas$, and $\Lgap$. In our model, each of these parameters has a halo mass dependence and redshift dependence. As the median redshifts of our samples are similar, we neglect the redshift dependence in this study. Also given the relatively narrow luminosity range of our samples, we only keep the halo mass dependence for $\Ac$, $\Lc$, and $\phis$, assuming they all have a power-law dependence on halo mass. Each of them is parameterised by the value at a pivot mass (chosen to be $10^{11}\hinvMsun$; with the corresponding parameter labelled with a subscript `p') and the power-law index of the mass dependence. In total, we have 9 parameters to simultaneously model the clustering measurements of all the luminosity-bin samples, which are $\Acp$, $\gammaA \equiv {\rm d}\log \Ac/{\rm d}\log \Mh$, $\Lcp$, $\gammaL \equiv {\rm d}\log \Lc/{\rm d}\log \Mh$, $\sigc$, $\phisp$, $\gammaphi \equiv {\rm d}\log \phis/{\rm d}\log \Mh$, $\alphas$, and $\Lgap$.

Given a set of the 9 CLF parameters and the luminosity cuts ($L_1$, $L_2$) of a galaxy sample, we compute the mean occupation functions of central and satellite galaxies, using equations~(\ref{eqn:Ncen}) and (\ref{eqn:Nsat}). Then, with the mean occupation functions, we can calculate the model number density
\begin{equation}
    \numden = \int_0^\infty \langle N(\Mh)\rangle \frac{{\rm d}n}{{\rm d}\Mh} {\rm d}\Mh,
\end{equation}
with $\langle N(\Mh)\rangle = \langle N_{\rm cen}(\Mh)\rangle +  \langle N_{\rm sat}(\Mh)\rangle$ and ${\rm d}n/{\rm d}\Mh$ the halo mass function. We follow the method developed in \citet{Zheng04} and improved in \citet{Tinker05} to calculate the 2PCF of the galaxy sample. In our work, we adopt the halo definition that the mean density of haloes is 200 times that of the background universe. The number of satellite galaxies in haloes of fixed mass is assumed to follow the Poisson distribution with the mean given by equation~(\ref{eqn:Nsat}). The spatial distribution of satellite galaxies within haloes is assumed to be the same as that of the dark matter, following the Navarro-Frenk-White profile \citep{NFW96}. The concentration parameter of the profile, $c(\Mh) = [c_0/(1+z)](\Mh/M_{\rm nl})^\beta$, with $c_0=11$, $\beta=-0.13$, and $M_{\rm nl}=3.79\times 10^{12}\hinvMsun$ (the nonlinear mass scale at $z=0$ for the adopted cosmology). 

\begin{figure}
  \centering
  \includegraphics[width=.5\textwidth]{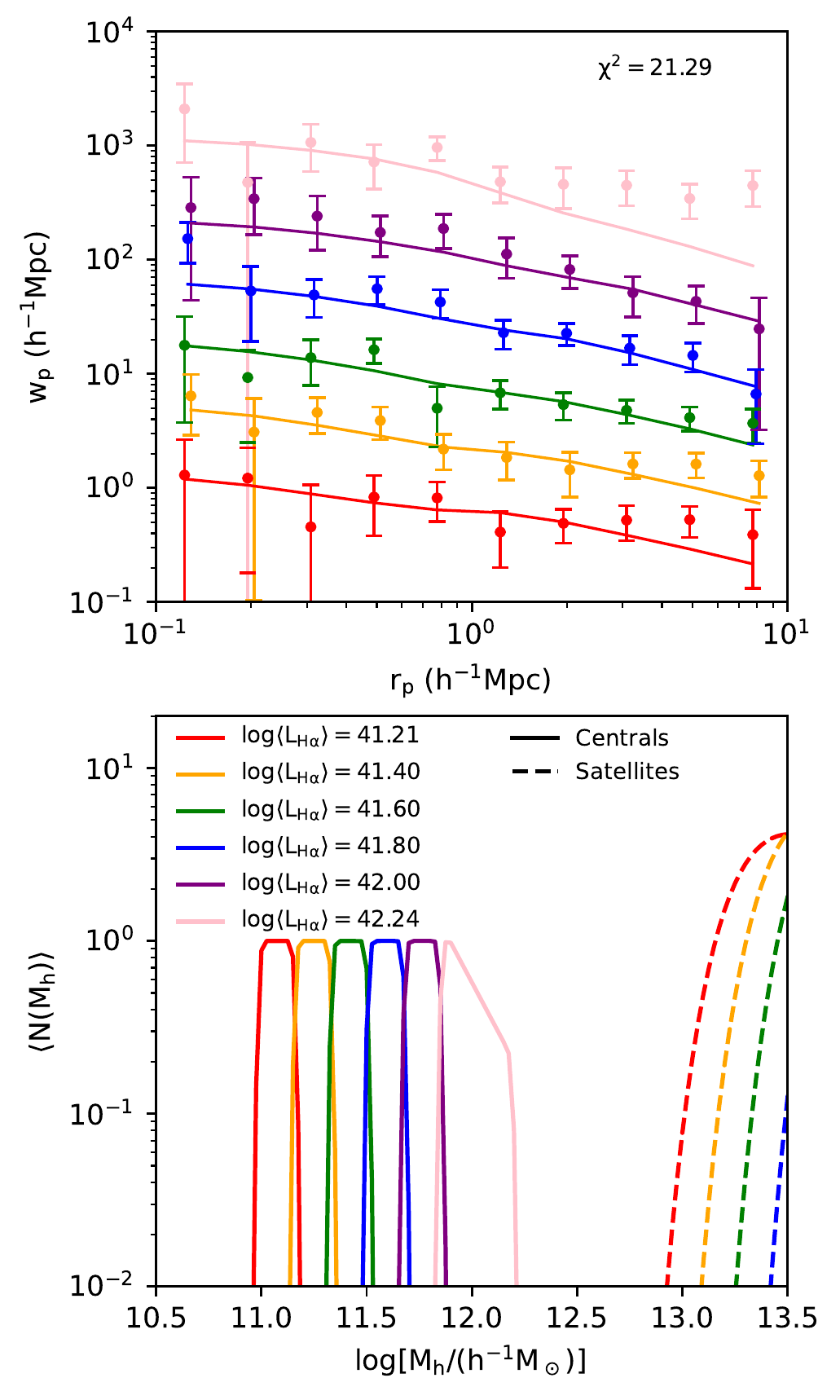}
  \caption{Illustration of halo modelling results for the luminosity dependent clustering. 
  Top: projected 2PCF measurements (points) and those from the best-fitting model (curves), colour-coded by \Ha\ luminosity. For clarity, the 2PCFs are each staggered by 0.5 dex. The model reproduces the observed clustering with good accuracy.
  Bottom: mean occupation function of each sample, inferred from the best-fitting model. The bump represents that of central galaxies, and the power-law-like curve is that of satellite galaxies. Galaxies with increasing \Ha\ luminosity live in dark matter halos of increasing mass. 
  }
  \label{fig:hod_lum_bins}
\end{figure}

\subsubsection{Model for the Luminosity and Stellar Mass Dependent Clustering}
\label{sec:CMLD}

To model the clustering of luminosity-bin samples with a cut in stellar mass, we adopt a formalism similar to that in \citet{Xu18}. To model the dependence of galaxy clustering on luminosity and colour, \citet{Xu18} develop a global parameterisation of the colour and luminosity distribution as a function of halo mass, namely the conditional colour-magnitude distribution (CCMD), and jointly model the clustering of a large number of galaxy samples defined by cuts in colour and luminosity. In this work, we parameterise the joint distribution of stellar mass and \Ha\ luminosity as a function of halo mass, separated into that for central and that for satellite galaxies. This extends the CLF by adding one more dimension.

For the ease of presenting the formalism, we use `$x$' to represent the logarithmic of stellar mass, $x\equiv\log M_*$, and `$y$' the logarithmic of \Ha\ luminosity, $y\equiv\log L_{\rm H\alpha}$. For central galaxies, the stellar mass -- \Ha\ luminosity distribution inside haloes of fixed mass, i.e. the conditional $M_*$--$L_{\rm H\alpha}$ distribution, is parameterised as a 2D Gaussian distribution, 
\begin{equation}
\frac{{\rm d}^2 \langle N_{\rm cen}(\Mh) \rangle}{{\rm d}x\, {\rm d}y} =
\frac{A_{\rm c}}{2\pi\sigma_x\sigma_y
\sqrt{1-\rho^2}}\exp\left[-\frac{Z^2}{2(1-\rho^2)}\right],
\label{eqn:dNcen_dxdy}
\end{equation}
with
\begin{equation}
Z^2 =  \frac{(x-\mu_x)^2}{\sigma^2_x}+\frac{(y-\mu_y)^2}
 {\sigma^2_y} - \frac{2\rho(x-\mu_x)(y-\mu_y)}
 {\sigma_x\sigma_y}
\end{equation}
and
\begin{equation}
\rho=\frac{{\rm Cov}(x,y)}{\sigma_x\sigma_y}.
\end{equation}
Here $A_{\rm c}$ is the fraction of haloes at the given mass occupied by star-forming galaxies, $\mu_x$ and $\mu_y$ are the mean logarithmic stellar mass and \Ha\ luminosity in these haloes, $\sigma_x$ and $\sigma_y$ are the corresponding standard deviations, and $\rho$ (${\rm Cov}(x,y)$) is the coefficient (covariance) of the correlation between logarithmic stellar mass and \Ha\ luminosity. While all these parameters can depend on halo mass, we find that some of them are not well constrained. Given the limited range of halo mass expected for the samples we consider, we assume no halo mass dependence for $\sigma_x$ and $\sigma_y$ and fix $\rho=0$.

The occupation fraction is assumed to have a power-law dependence on halo mass. That is,
\begin{equation}
    \log A_{\rm c}=\log A_{\rm c,p} + \gamma_A (\log\Mh-\log M_{\rm h,p}),
\end{equation}
where $M_{\rm h,p}=10^{11}\hinvMsun$ is the pivot halo mass and the quantity $A_{\rm c,p}$ is the value at the pivot halo mass. In this work, we neglect the halo mass dependence and fix $\gamma_A=0$. For the halo mass dependence of the median stellar mass and \Ha\ luminosity, we follow \citet{Xu18} to parameterise each to be a power law with exponential cutoff at low halo mass. As an example, the median stellar mass has the following form
\begin{equation}
    M_{\rm *,m} = M_{\rm *,t} \left(\frac{\Mh}{\Mtx}\right)^{\alpha_x}\exp\left[-\frac{\Mtx}{\Mh}+1\right],
\end{equation}
where $\Mtx$ denotes a transition halo mass scale, $M_{\rm *,t}$ is the median stellar mass at this halo mass scale, and $\alpha_x$ is the power-law index. It reduces to 
\begin{equation}
\mu_x = \mu_{xt} + \alpha_x(\log\Mh-\log\Mtx)+(-\Mtx/\Mh+1)/\ln 10,
\label{eqn:mux}
\end{equation}
with $\mu_{xt}=\log M_{\rm *,t}$. Similarly, for the median \Ha\ luminosity, we have the form
\begin{equation}
\mu_y = \mu_{yt} + \alpha_y(\log\Mh-\log\Mty)+(-\Mty/\Mh+1)/\ln 10.
\label{eqn:muy}
\end{equation}

In total, there are nine free parameters for the conditional $M_*$--$L_{\rm H\alpha}$ distribution of central galaxies, $A_{\rm c,p}$, $\mu_{xt}$, $\Mtx$, $\alpha_x$, $\mu_{yt}$, $\Mty$, $\alpha_y$, $\sigma_x$, and $\sigma_y$.

For satellite galaxies, the conditional stellar mass function (stellar mass function at fixed halo mass) is described by a modified Schechter function \citep[e.g.][]{Yang08,Xu18},
\begin{equation}
    \frac{{\rm d} \langle N_{\rm sat}(\Mh) \rangle}{{\rm d} \log M_*} =
\phi_{\rm s} \left( \frac{M_*}{M_{*,{\rm s}}}
\right)^{\alpha_{\rm s} + 1} \exp \left[ -\left( \frac{M_*}{M_{*,{\rm s}}} \right)^2 \right],
\end{equation}
where  $\phi_{\rm s}$ is the normalisation, $\alpha_{\rm s}$ is the faint-end slope, and $M_{*,{\rm s}}$ is the characteristic stellar mass.
To obtain the paramerisation of the conditional $M_*$--$L_{\rm H\alpha}$ distribution for satellites, we adopt the above form for the stellar mass distribution and assume that (the logarithmic) \Ha\ luminosity follows a Gaussian distribution at fixed stellar mass \citep{Xu18},
\begin{multline}
\frac{{\rm d}^2 \langle N_{\rm sat}(\Mh) \rangle}{{\rm d}x\, {\rm d}y}
 =
 \frac{1 }{\sqrt{2\pi}\sigma_{y,{\rm sat}}} \exp\left[- \frac{
 (y-\mu_{y,{\rm sat}})^2}{2\sigma^2_{y,{\rm sat}}} \right]  \\
 \times \phi_{\rm s}
10^{(\alpha_{\rm s}+1)(x-x_{\rm s})}
\exp \left[ -10^{2(x-x_{\rm s})}  \right],
\end{multline}
where $x_{\rm s}\equiv \log M_{*,{\rm s}}$. We parameterise $x_{\rm s}$ through the stellar mass gap, the difference between the (logarithmic) central and satellite characteristic stellar mass,
\begin{equation}
    x_{\rm s}=\mu_x - \Delta_{\rm cs}.
\end{equation}
The amplitude $\phi_{\rm s}$ has a dependence on halo mass
\begin{equation}
    \log \phi_{\rm s}=\log \phi_{\rm s,p}+\gamma_\phi (\log\Mh-\log M_{\rm h,p}).
\end{equation}
The median \Ha\ luminosity has a dependence on stellar mass, motivated by the relation on the star formation main sequence,
\begin{equation}
    \mu_{y,{\rm sat}}=\mu_{yp,{\rm sat}}+\gamma_{ys} (x-x_p),
\end{equation}
where the pivot stellar mass is taken to be $10^{10}\Msun$, i.e. $x_p=10$. 

In the above, we choose to present the overall framework so that it can be applied to future surveys. In this work, given the sample size and the uncertainties in the measurements, the satellite occupation distribution cannot be tightly constrained. Therefore, we apply strong priors broadly motivated by previous work. For example, we set the gap parameter $\Delta_{\rm cs}$ to zero and require the conditional stellar mass function of satellites to be higher than 0.1 at $\log (M_*/\Msun)=10$ in $\log[\Mh/(\hinvMsun)]=12$ haloes \cite[e.g.][]{Leauthaud12,Lim17}. We also require the satellite star formation main sequence to be close to the central one, with $\mu_{y, {\rm sat}}$ to be within $0.5 \sigma_{y, {\rm cen}}$ of $\mu_{y, {\rm cen}}$ in haloes of $\log[\Mh/(\hinvMsun)]=12$. With much larger samples from future surveys, such priors would not be necessary and the parameters would be constrained by the data. In total, we have six parameters to describe the conditional $M_*$--$L_{\rm H\alpha}$ distribution of satellite galaxies, $\phi_{\rm s,p}$, $\gamma_\phi$, $\alpha_{\rm s}$, $\mu_{yp,{\rm sat}}$, $\gamma_{ys}$, and $\sigma_{y,{\rm sat}}$.

With the above parameterisation of the conditional $M_*$--$L_{\rm H\alpha}$ distribution, for a sample defined by cuts in stellar mass and \Ha\ luminosity, we can integrate the distribution to obtain the mean occupation functions of central and satellite galaxies for this sample, which can then be used to compute the model 2PCF. As in \citet{Xu18}, this global parameterisation allows us to simultaneously model the 2PCFs of a number of galaxy samples constructed with different stellar mass and \Ha\ luminosity cuts.

\bigskip
\bigskip

With the above CLF (Section~\ref{sec:CLF}) or conditional $M_*$--$L_{\rm H\alpha}$ distribution (Section~\ref{sec:CMLD}) model setups, for a given set of parameters, we calculate the model prediction for the projected 2PCF $w_p$ and galaxy number density for each galaxy sample and form the $\chi^2$ as
\begin{equation}
    \chi^2 = \mathbf{ (w_p - w_p^*)^{\mathbf{T}} {\rm C}^{-1} (w_p - w_p^*)} + 
    \mathbf{ (n_g - n_g^*)^{\mathbf{T}} {\rm C_n}^{-1} (n_g - n_g^*) },
\end{equation}
where $\mathbf{w_p}$ and $\mathbf{n_g}$ are the vectors of the projected 2PCFs and number densities for either all the luminosity-dependent samples or all the luminosity-and-stellar-mass-dependent samples with the measured values denoted with a superscript `*', $\mathbf{\rm C}$ is the full covariance matrix of the projected 2PCFs, and $\mathbf{\rm C_n}$ is the covariance matrix of the number densities. As we have limited area to estimate the 2PCF covariance matrix of each sample using the jackknife method, we neglect the covariance between different samples. For the number density covariance matrix, a 10\% uncertainty is assumed for each sample. The final covairance matrix is scaled by $(N-N_{\rm d}-2)/(N-1)$ to account for the mean bias in inverting the matrix \citep{Hartlap07}, where $N=128$ is the number of jackknife samples and $N_{\rm d}$ is the number of data points. With the above $\chi^2$, the likelihood of the model is proportional to $\exp(-\chi^2/2)$, and a Markov Chain Monte Carlo (MCMC) method is employed to explore the parameter space and to obtain constraints on the model. 

\begin{figure*}
  \includegraphics[width=1.0\textwidth]{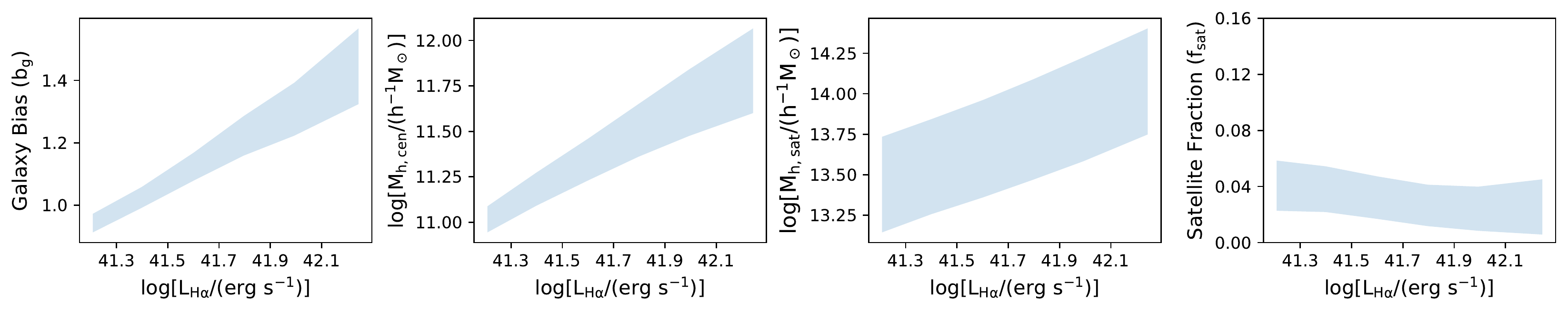}
  \caption{Derived quantities from modelling the \Ha\ luminosity dependent clustering. All quantities are shown as a function of \Ha\ luminosity of galaxies.
  From left to right are the large-scale galaxy bias factor, the median mass of haloes hosting central galaxies, the median mass of haloes hosting satellite galaxies, and the satellite fraction. In each panel, 
  the shaded region shows the central 68.3 per cent of the distribution. The trends of galaxy bias and $M_{\rm h, cen}$ increasing \Ha\ luminosity are apparent while the luminosity dependence of $M_{\rm h,sat}$ and the satellite fraction are very weak.
  }
  \label{fig:derived_lum_bins}
\end{figure*}

\subsection{Modelling Results for the Luminosity Dependent Clustering}
\label{sec:CLF_model_results}

The constraints on the CLF parameters from simultaneously modelling all the luminosity-bin samples are shown in Fig.~\ref{fig:corner_lum_bins}. In this subsection we focus on the halo occupation functions and several derived quantities.

In the top panel of Fig.~\ref{fig:hod_lum_bins}, the projected 2PCFs (curves) from the best-fitting model are plotted with the measurements (points). For clarity, offsets are added to both the model curves and data points. The model provides a good description of the measurements, with $\chi^2\simeq 21.3$ for 57 degrees of freedom (10 $w_p$ points and 1 number density for each of the six luminosity-bin samples, minus 9 free parameters). 

The mean occupation functions derived from the best-fitting model are shown in the bottom panel of Fig.~\ref{fig:hod_lum_bins}. For each sample, the bump at low mass is the mean occupation function of central galaxies. Roughly speaking, central galaxies in our samples reside in haloes of masses $\sim 10^{11-12}\hinvMsun$. In this best-fitting model, nearly all the halos in this mass range are occupied by star-forming galaxies with \Ha\ emission. In fact, from the parameter constraints in Fig.~\ref{fig:corner_lum_bins}, the occupation fraction is mostly constrained to be order unity. The amplitude parameter of the central galaxy CLF, $A_{\rm c,p}$ at the pivot mass $10^{11}\hinvMsun$ is about unity or larger than unity. Together with the mass dependence parameters, $\gamma_A\equiv {\rm d}\log A_{\rm c}/{\rm d}\log M_h$, we can derive the occupation fraction $A_{\rm c}$ of central galaxies in haloes of any given mass. When this fraction is larger than unity, we set it to be unity in our model. Therefore, the preferred larger-than-unity value of $A_{\rm c,p}$ means unity occupation of central galaxies in low mass haloes, and the data is consistent with such a nearly unity occupation. Only in high mass haloes, the occupation fraction starts to drop. 

We note that in Fig.~\ref{fig:hod_lum_bins} the mean central galaxy occupation function (the bump) for each sample appears to have relatively sharp edges, resulting from the small scatter ($\sigma_c$) in stellar mass at fixed halo mass for the best-fitting HOD used for the illustration. In fact, $\sigma_c$ is loosely constrained, with the 2$\sigma$ range being $\sim 0$--$0.49$ dex (Fig.~\ref{fig:corner_lum_bins}), which is not reflected in the illustration. The overall trend of the dependence of the occupation on \Ha\ luminosity and halo mass, however, is not affected by such uncertainties.

In the bottom panel of Fig.~\ref{fig:hod_lum_bins}, the mean occupation function for satellite galaxies in each sample can be approximately described by a steep power law. It suggests that star-forming satellite galaxies can be found in massive halos. The constraints mainly come from the small-scale clustering, in the one-halo regime. If there were no satellite galaxies, there would be no inter-halo galaxy pairs and the real-space 2PCF would drop to zero below Mpc scale. As a consequence, the projected 2PCF $w_p$ would become flattened below such a scale. The data tend to have a $w_p$ profile increasing toward the smallest scale and a fraction of galaxies can be satellites. From parameter constraints in Fig.~\ref{fig:corner_lum_bins}, we can see that the overall constraints on satellite occupation function is loose --- the amplitude $\phisp$ at the pivot mass $10^{11}\hinvMsun$ and the luminosity gap $\Lgap$ both vary by more than one dex.

The posteriors for derived quantities can be inferred from the MCMC chain. In Fig.~\ref{fig:derived_lum_bins}, the constraints on four derived quantities are shown. The left-most panel is the large-scale galaxy bias factor $b_g$ as a function of \Ha\ luminosity, where $b_g$ is the halo bias $b_h$ properly weighted by the mean occupation and halo mass function,
\begin{equation}
    \bg = \frac{1}{\numden}\int_0^\infty b_h(\Mh)\langle N(\Mh)\rangle \frac{{\rm d}n}{{\rm d}\Mh} {\rm d}\Mh.
\end{equation}
Approximately, the galaxy bias factor appears to increase linearly with $\log L_{\rm H\alpha}$. As the real-space 2PCF scales as $\bg^2$, under the power-law approximation we can connect $\bg$ to the clustering strength $r_0$, $\bg\propto r_0^{\gamma/2}$. Our derived $\bg$ trend is consistent with the $r_0$ trend seen in the left panel of Fig.~\ref{fig:r0_lum}. 
 
The second panel in Fig.~\ref{fig:derived_lum_bins} shows the luminosity dependence of the median mass of haloes hosting central galaxies. The median mass $M_{\rm h,cen}$ is derived through
\begin{equation}
    \int_0^{M_{\rm h,cen}} \langle N_{\rm cen}(\Mh)\rangle \frac{{\rm d}n}{{\rm d}\Mh} {\rm d}\Mh = \frac{1}{2} \int_0^\infty \langle N_{\rm cen}(\Mh)\rangle \frac{{\rm d}n}{{\rm d}\Mh} {\rm d}\Mh .
\end{equation}
From the dependence, the implied halo mass-\Ha\ luminosity relation of central galaxies approximately follows a power law, $\Mh \propto L_{\rm H\alpha}^{0.75}$. The third panel shows the median mass of haloes hosing satellite galaxies, calculated similarly. It is around $10^{13.5-14.0}\hinvMsun$, but the constraint is not as tight as that for central galaxies, with an uncertainty about 0.5--0.8 dex, showing only a weak luminosity dependence.

The right-most panel in Fig.~\ref{fig:derived_lum_bins} plots the luminosity dependent fraction of galaxies being satellites.
As with the median mass of satellite hosting haloes, the constraint on the satellite fraction is loose (from nearly zero per cent to about five per cent), and it is consistent with no dependence on \Ha\ luminosity.

\subsection{Modelling Results for the Luminosity and Stellar Mass Dependent Clustering}
\label{sec:HOD_lum_mass}

\begin{figure*}
  \centering
  \includegraphics[width=1.0\textwidth]{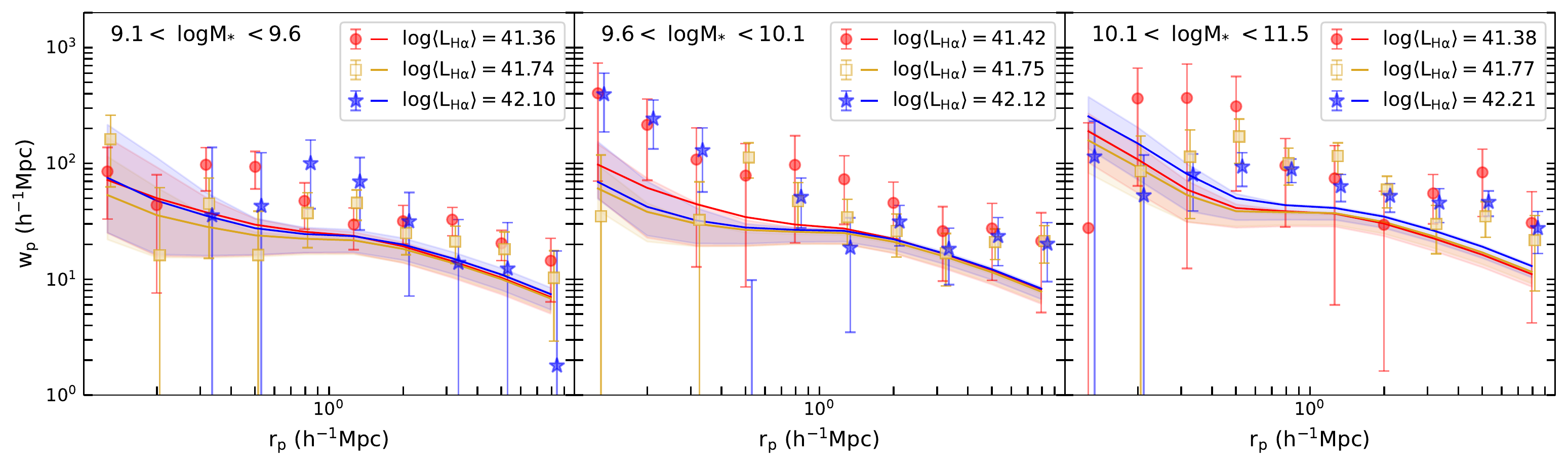}
  \caption{
  Measurements (points) of the stellar mass and \Ha\ luminosity dependent projected 2PCFs and fits (curves) from the conditional $M_*$--$L_{\rm H\alpha}$ distribution model. The three panels are for three different stellar mass bins. In each panel the dependence of the 2PCF on \Ha\ luminosity is shown, and the measurements from different \Ha\ luminosity samples are shifted slightly in the horizontal direction for clarify. The clustering shows a clear trend that the amplitude increases with increasing stellar mass, but at fixed stellar mass there is only a weak dependence on \Ha\ luminosity. Although the model predicts slightly lower amplitudes, it provides good fits ($\chi^2\simeq 22.1$ for 84 degrees of freedom) and reproduces the major trends. 
}
  \label{fig:pred_wp_mass_bins}
\end{figure*}

\begin{figure}
  \includegraphics[width=.45\textwidth]{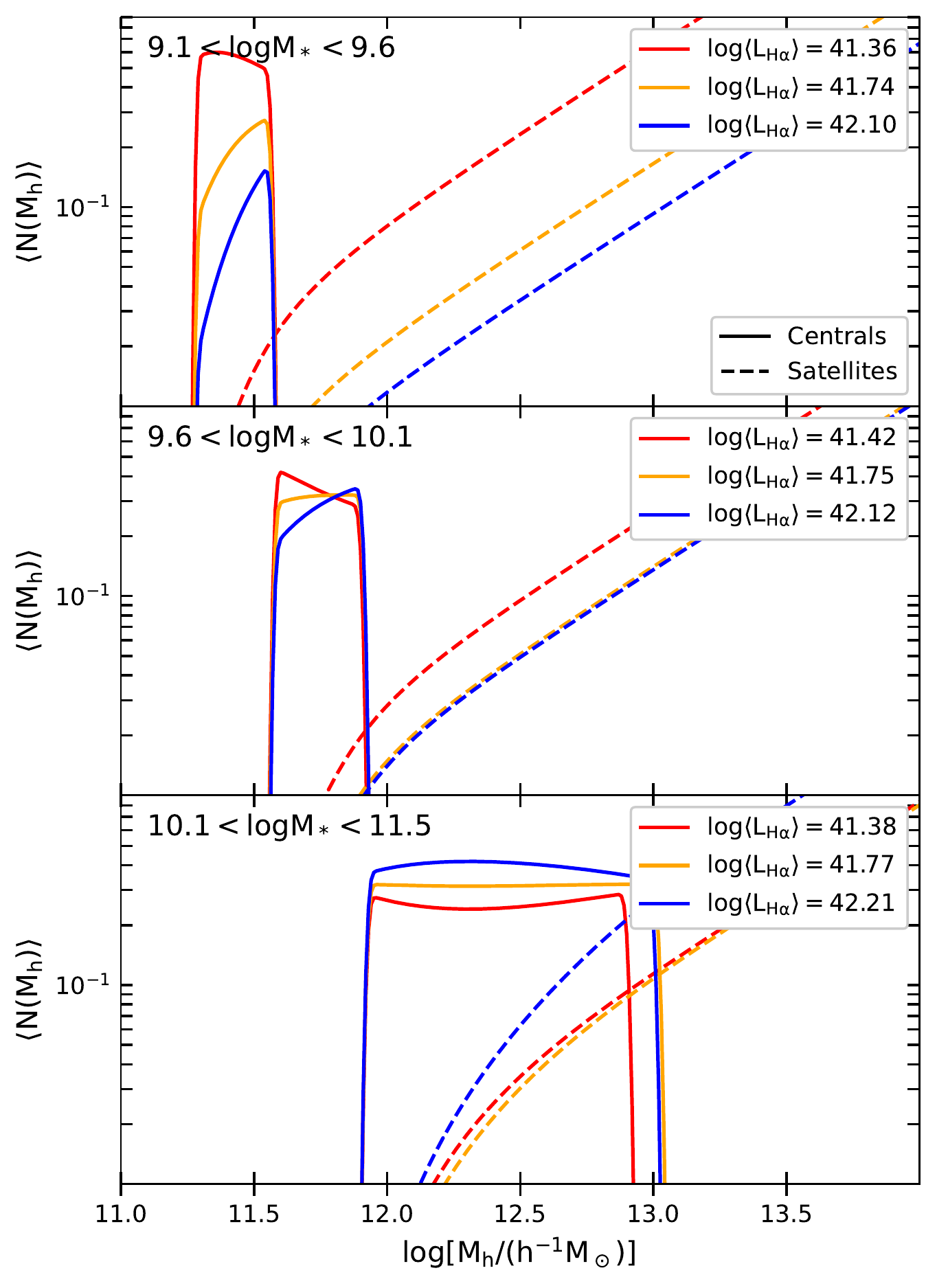}
  \caption{
  Illustration of the mean occupation function of central galaxies (solid) and satellite galaxies (dashed) as a function of stellar mass and \Ha\ luminosity. We show the prediction from the best-fitting model from jointly modelling the stellar mass and \Ha\ luminosity dependent galaxy clustering based on the conditional $M_*$--$L_{\rm H\alpha}$ distribution formalism. From top to bottom panels, the stellar mass of the samples increases, and in each panel colour-coded is the \Ha\ luminosity. Here stellar mass of central galaxies show a tighter correlation with halo mass than \Ha\ luminosity.
  }
  \label{fig:nsat_stellar_mass_bins}
\end{figure}

The formalism of the conditional $M_*$--$L_{\rm H\alpha}$ distribution (Section~\ref{sec:CMLD}) can be applied to model the clustering of galaxies samples defined by arbitrary cuts in $M_*$ and $L_{\rm H\alpha}$. To have samples across a large range in stellar mass and \Ha\ luminosity while retaining a reasonable signal-to-noise ratio, we apply cuts to create three stellar mass bins and in each stellar mass bin to construct three \Ha-luminosity dependent samples. The $\log(M_*/\Msun)$ range for the three stellar mass bins are  $9.1 < \log M_* < 9.6$, $9.6 < \log M_* < 10.1$, and $10.1 < \log M_* < 11.5$, respectively. In each stellar mass bin, the three \Ha-luminosity dependent samples have the following ranges of $\log[L_{\rm H\alpha}/({\rm erg\, s^{-1}})]$: $\log L_{\rm H\alpha}<41.6$, $41.6<\log L_{\rm H\alpha} <41.9$, and $\log L_{\rm H\alpha}>41.9$. See the last set of samples in Table~\ref{tab:mass_bin_lum_sample}.

The $w_p$ measurements and the number densities of the nine samples are modelled simultaneously with the conditional $M_*$--$L_{\rm H\alpha}$ distribution formalism. Fig.~\ref{fig:pred_wp_mass_bins} shows the measurements and the model fits. While the model fits appear to be slightly lower in amplitude than the measurements, given the uncertainties in the measurements, the model provides reasonable fits, $\chi^2=22.1$ for 84 degrees of freedom (99 data points and minus 15 model parameters). The measured 2PCFs show a clear dependence on stellar mass, with stronger clustering for samples of higher stellar mass. For galaxies in each stellar mass bin, the measured 2PCFs show little dependence on \Ha\ luminosity. These trends are well captured by the model.

In Fig.~\ref{fig:nsat_stellar_mass_bins}, we illustrate the mean occupation functions of central and satellite galaxies in these samples, derived from the best-fitting model of the conditional $M_*$--$L_{\rm H\alpha}$ distribution. In a given stellar mass bin, central galaxies of different \Ha\ luminosity occupy haloes in a similar mass range. As stellar mass increases, the mass scale of hosting haloes also increases. That is, the model shows a tight correlation between central galaxy stellar mass and halo mass, while the correlation between \Ha\ luminosity and halo mass is relatively weak. This can also be seen from the scatter in the central stellar mass and \Ha\ luminosity at fixed halo mass, represented by the parameters $\sigma_x$ and $\sigma_y$ in equation~(\ref{eqn:dNcen_dxdy}), with the median value for the former at the level of $\sigma_x\sim 0.03$ and the latter at the level of $\sigma_y\sim 0.35$ (see Fig.~\ref{fig:corner_m*}). For each sample, the mean satellite occupation function approximately follows a power law, with the satellite fraction around 10--20 per cent but not well constrained (see Table~\ref{tab:derived M_*}). As the 2PCFs on large scales are dominated by contributions from central galaxies, the tight correlation between stellar mass and halo mass explains the main trends seen in Fig.~\ref{fig:pred_wp_mass_bins}.

As with Fig.~\ref{fig:hod_lum_bins}, we caution that the uncertainties in the HOD constraints are not reflected in Fig.~\ref{fig:nsat_stellar_mass_bins}, which, when considered, would lead to less sharp edges in the mean occupation functions for central galaxies. Nevertheless, the HOD modelling enables us to infer the overall trend of the occupation with stellar mass and \Ha\ luminosity.

\begin{figure}
  \includegraphics[width=.375\paperwidth]{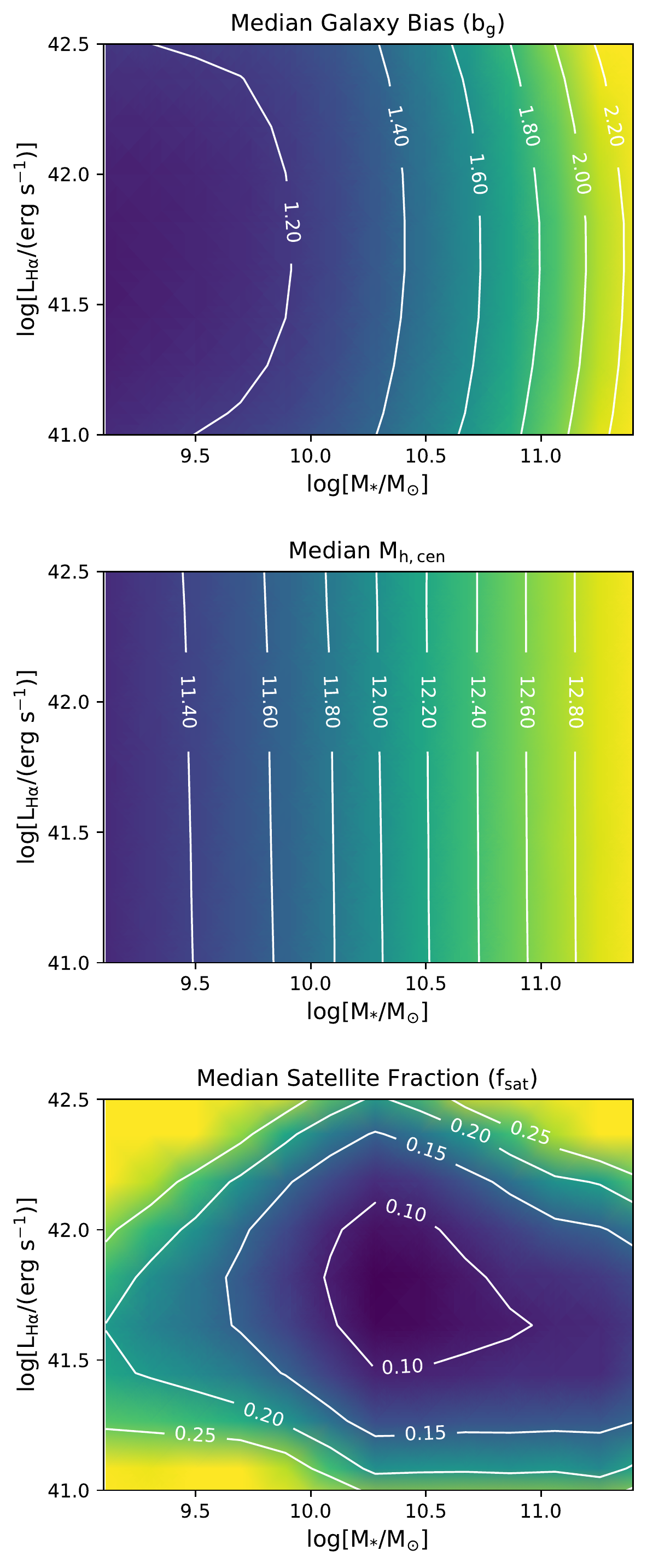}
  \caption{
  Median galaxy bias (top), median mass of haloes hosting central galaxies (middle), and median satellite fraction (bottom), as a function of galaxy stellar mass and \Ha\ luminosity, derived based on the chains from jointly modelling stellar mass and \Ha\ luminosity dependent galaxy clustering based on the conditional $M_*$--$L_{\rm H\alpha}$ distribution formalism. In the top panel, except for the lowest stellar mass, the contours are more vertical than horizontal, implying stronger dependence on stellar mass than on \Ha\ luminosity. In the middle panel, the contours are nearly vertical, indicating that for central galaxies stellar mass is more correlated with halo mass than \Ha\ luminosity. The bottom panel shows that at fixed stellar mass satellite fraction is high at both low and high \Ha\ luminosity, which explains the curvature seen in the contours of galaxy bias in the top panel (as well as in the clustering amplitude in the right panel of Fig.~\ref{fig:r0_stellar_mass}).
  }
  \label{fig:median_bias_median_mcen}
\end{figure}

To further examine the implied correlation of stellar mass and \Ha\ luminosity with halo mass from the model, in Fig.~\ref{fig:median_bias_median_mcen} we show the median galaxy bias and the median halo mass as a function of galaxy stellar mass and \Ha\ luminosity, inferred from all the models in the MCMC chain.  For galaxy bias (top panel), except for the low stellar mass end, the contours are more vertical than horizontal, implying that galaxy clustering has a stronger dependence on stellar mass than on \Ha\ luminosity. This is another representation of the trends seen in Fig.~\ref{fig:wp_fixed_sm}, Fig.~\ref{fig:r0_stellar_mass} (right panel), and Fig.~\ref{fig:pred_wp_mass_bins}. In the middle panel of Fig.~\ref{fig:median_bias_median_mcen}, the contours are nearly vertical, demonstrating that the tight correlation is between central galaxy stellar mass (rather than \Ha\ luminosity) and halo mass. 

Interestingly, although the contours in the top panel of Fig.~\ref{fig:median_bias_median_mcen} are much more vertical than horizontal, they do show a small amount of curvature. That means, at fixed stellar mass the galaxies with the lowest and highest \Ha\ luminosities have a slightly higher galaxy bias. This trend is entirely consistent with the trends shown in the right panel of Fig.~\ref{fig:r0_stellar_mass}, where $r_0$ shows a V-shaped dependence on \Ha\ luminosity. Since there is virtually no such curvature for central galaxy median halo mass seen in the middle panel of Fig.~\ref{fig:median_bias_median_mcen}, we can conclude that the dependence of bias on \Ha\ luminosity at fixed stellar mass is most likely caused by the satellite galaxy population. If at fixed stellar mass satellite galaxies had a broader distribution in \Ha\ luminosity, and hence sSFR, than equivalent centrals, we would expect just such a result. Indeed, with our parameterisation, the modelling results show such a trend -- from Fig.~\ref{fig:corner_m*}, we have the scatter in \Ha\ luminosity  $\sigma_y\sim 0.35$ dex (central galaxies) and $\sigma_{y,{\rm sat}}\sim 0.63$ dex (satellites). Further support comes from the bottom panel of Fig.~\ref{fig:median_bias_median_mcen}, which shows the satellite fraction in bins of stellar mass and \Ha\ luminosity from the model fit. At fixed stellar mass, the satellite fraction increases at both high and low \Ha\ luminosity, as shown in the bottom panel (also see Table~\ref{tab:derived M_*}). Since satellites are typically hosted by more massive halos than centrals of similar stellar mass, the clustering amplitude (hence galaxy bias) correspondingly increases. As expected, the HOD modelling provides a more informative interpretation of the clustering trend than can be inferred from the power-law fits alone. We caution the reader that the uncertainties on the best fit HOD parameters (Fig.~\ref{fig:corner_m*}) for the satellites, and hence on the satellite fractions (Table~\ref{tab:derived M_*}), are large and that the median values of the derived satellite fractions vary significantly between different HOD model formalisms (Tables~\ref{tab:derived} and \ref{tab:derived M_*}), so these trends are tentative at best. We highlight them here only because they are consistent with, and provide an explanation for, the V-shaped dependence of $r_0$ on \Ha\ luminosity at fixed stellar mass that we observe.

\begin{figure}
  \includegraphics[width=.4\paperwidth]{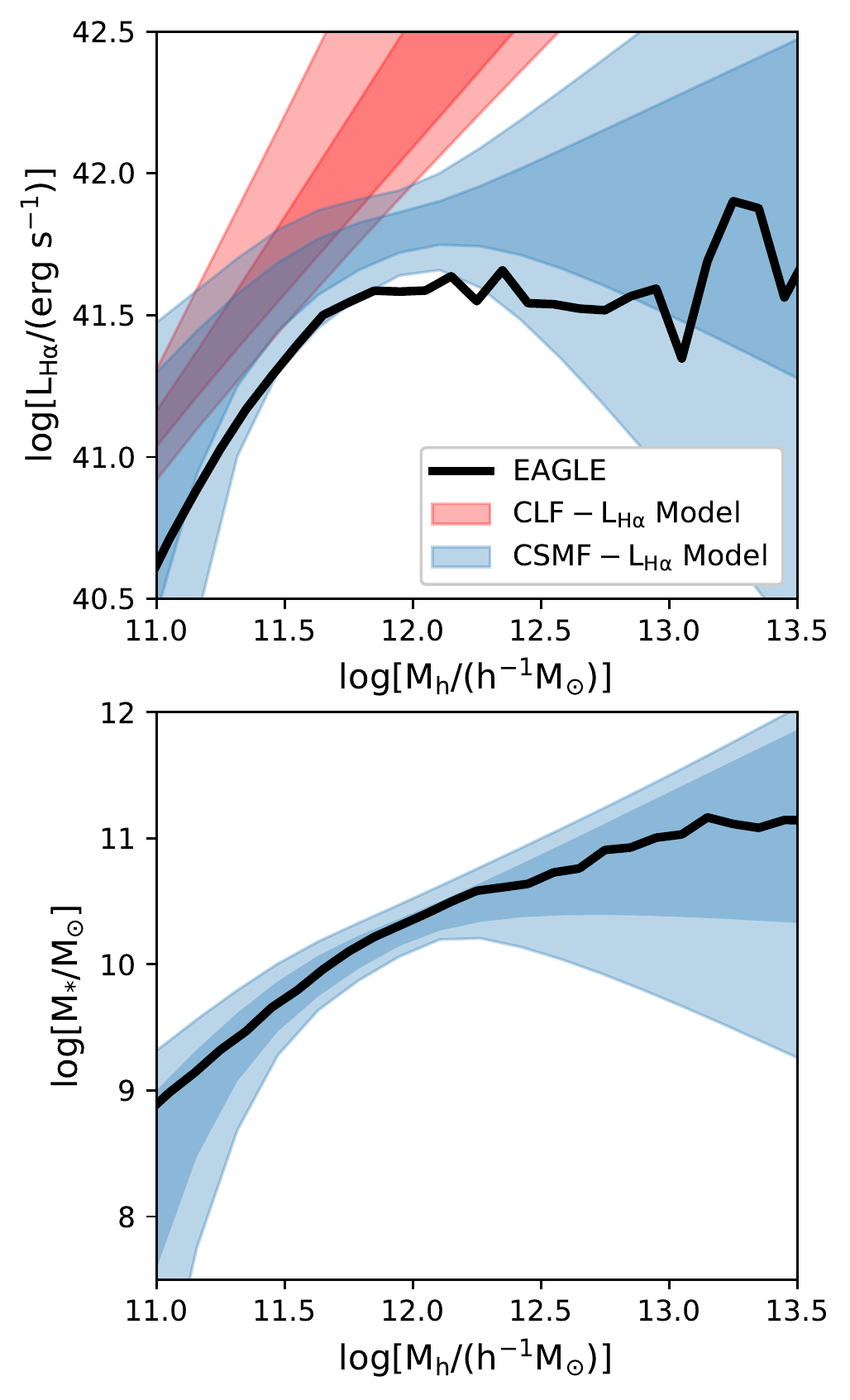}
  \caption{
  Relation between median \Ha\ luminosity of central galaxies and halo mass (top) and that between median stellar mass of central galaxies and halo mass (bottom). The blue shaded regions are the 1$\sigma$ and 2$\sigma$ constraints from modelling the stellar mass and \Ha\ luminosity dependent clustering with the conditional $M_*$--$L_{\rm H\alpha}$ distribution model, while the red shaded regions in the top panel are those from modelling the \Ha\ luminosity dependent clustering with the CLF model. In each panel, the solid curve is the relation predicted by the EAGLE hydrodynamic galaxy formation model. See the text for detail.
  }
  \label{fig:mstar_mh_lha_mh}
\end{figure}

In Fig.~\ref{fig:mstar_mh_lha_mh}, we show the median stellar mass (bottom) and median \Ha\ luminosity (top) of central galaxies as a function of halo mass from the model, following equations~(\ref{eqn:mux}) and (\ref{eqn:muy}). Both are parameterised as a power law toward the high halo mass end and an exponential cutoff toward the low halo mass end. The data (2PCFs and galaxy number densities) lead to good constraints on both relations in haloes around $10^{12}\hinvMsun$. The constraints become loose at the low and high mass end, as a result of sample limitations. There is a tendency that central \Ha\ luminosity in high mass haloes levels off, indicating less active star formation in central galaxies of high mass haloes. However, the loose constraints prevent us from drawing any robust conclusion.

The solid black curves in Fig.~\ref{fig:mstar_mh_lha_mh} are the predicted relations from the EAGLE hydrodynamic galaxy formation simulation with a box size of 100 Mpc (comoving), i.e. the run named Ref-L0100N1504 \citep[][]{Schaye15,EAGLE16}. For the comparison, we use the halo mass in the EAGLE simulation so that haloes have a mean density 200 times that of the background universe, consistent with what we adopt in this work. We convert the SFR in the simulation to \Ha\ luminosity using the relation in \citet{Kennicutt12} and then apply the stellar-mass-dependent dust extinction given by \citet{Garn10}. We note that both the EAGLE simulation and the 3D-HST stellar mass calculation adopt the \citet{Chabrier03} initial mass function (IMF). For the relation between median stellar mass and halo mass, the EAGLE curve falls right within the 1$\sigma$ range of our HOD model constraints. Since the EAGLE simulation calibrates its feedback parameters to reproduce the $z\sim 0.1$ stellar mass function, the excellent agreement in the $z\sim 1$ central galaxy stellar mass--halo mass relation with the HOD modelling result is encouraging. For the relation between central galaxy \Ha\ luminosity and halo mass, the EAGLE result has a trend similar to that in the HOD model constraint, but the curve lies slightly out of the 1$\sigma$ range of the model constraints. In haloes of $\sim 10^{12}\hinvMsun$, the EAGLE prediction is about 0.15 dex lower than the central constraint of the HOD model. This is in line with the result that the EAGLE simulation underpredicts the $z\sim 1$ SFR function at SFR$\sim$1--10$\Msun\, {\rm yr}^{-1}$ and the cosmic SFR density \citep{Katsianis17}, when compared with observationally derived values (e.g. from \Ha). The likely cause is that the supernova feedback adopted in the EAGLE simulation is too strong \citep{Katsianis17}. Overall, the HOD modelling results provide useful tests to the galaxy formation model. The broad agreement with the results from the EAGLE simulation, on the other hand, supports our model parameterisation.

The red shaded region in the top panel of Fig.~\ref{fig:mstar_mh_lha_mh} shows the constraints on the dependence of central galaxy \Ha\ luminosity on halo mass from the CLF-based model (Sections~\ref{sec:CLF} and \ref{sec:CLF_model_results}). At the low halo mass end, the CLF constraints agree with those from the conditional $M_*$--$L_{\rm H\alpha}$ distribution model. Although the uncertainty in the constraints becomes large  in high mass haloes, the CLF model tends to have a higher \Ha\ luminosity than the conditional $M_*$--$L_{\rm H\alpha}$  distribution model. We note that this is not necessarily a fair comparison -- we show here the luminosity only in halos hosting star-forming galaxies. Since the CLF model (red) has a relatively strong dependence of occupation fraction of star-forming galaxies on halo mass (e.g. with the power-law index of $\sim -2.09$; see Table~\ref{tab:CLFpar} and Fig.~\ref{fig:corner_lum_bins}), in high mass haloes (e.g. $\Mh\sim 10^{12.5}\hinvMsun$) there is a large fraction of haloes with no star-forming galaxies occupied at the centre. If we were to compare the \Ha\ luminosity averaged over all haloes (with or without central star-forming galaxies) of fixed mass, we expect to see a better overlap between the red and blue shaded region at the high halo mass end. In addition, the origin of the difference also lies in the difference in the model parameterisation. In the CLF model, central galaxy \Ha\ luminosity is assumed to have a power-law dependence on halo mass, while in the conditional $M_*$--$L_{\rm H\alpha}$ distribution model there is more flexibility to effectively allow the power-law index to change with halo mass. In principle, we could make the CLF model parametrisation more flexible. Since the CLF model already provides good fits to the \Ha\ luminosity dependent clustering, we leave it in its current form to show the model degeneracy. With data sets much larger than used here, however, a more flexible form is necessary and we also advocate modelling galaxy clustering within the formalism like the conditional $M_*$--$L_{\rm H\alpha}$ distribution to efficiently explore the galaxy-halo connection.

\section{Discussion}
\label{sec:discussion}

\begin{figure*}
  \includegraphics[width = \textwidth]{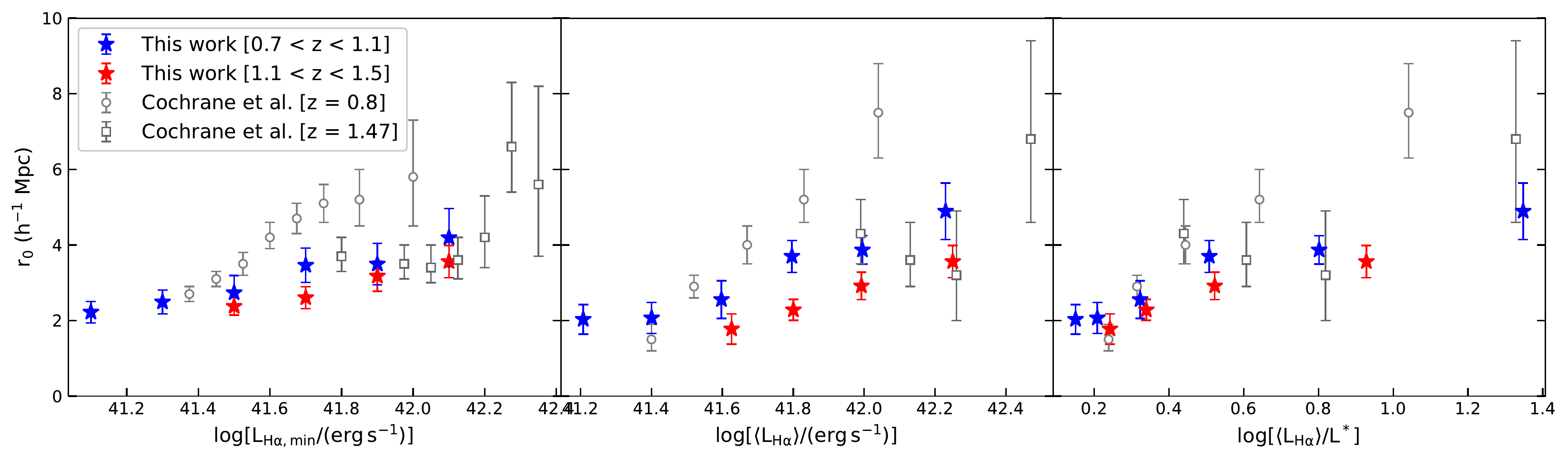}
  \caption{Comparison of the clustering strengths inferred from our work with those from \citet{cochrane17} for luminosity-threshold and luminosity-bin samples. Left: clustering strength as a function of \Ha\ luminosity threshold. Center: clustering strength as a function of mean \Ha\ luminosity. Right: clustering strength as a function of \Ha\ luminosity normalised by the characteristic luminosity $L^*$ of the \Ha\ luminosity function at each redshift. Our results agree within two sigma across the two samples though Cochrane et. al see higher $r_0$ values perhaps due to the differences in the  redshifts ranges of our samples. All samples are fitted with a fixed $\gamma$ of 1.8.
  }
  \label{fig:r0_vs_cochrane}
\end{figure*}

In this work, we construct samples of \Ha\ emitting galaxies with $0.7 < z < 1.5$ based on the grism spectra from the 3D-HST survey, study the \Ha\ luminosity and stellar mass dependent clustering, and model those clustering measurements using the CLF model and the conditional $M_*$--$L_{\rm H\alpha}$ distribution model within the HOD formalism. Looking both directly at the measurements and the resulting halo model fits, we see a number of clear results emerging. There is a strong dependence of the clustering amplitude of galaxies on their \Ha\ luminosity with more luminous galaxies being more strongly clustered. However, this trend is largely being driven by the correlation of \Ha\ luminosity with stellar mass and the well known relationship between stellar mass and clustering amplitude \citep[e.g.][]{wake,Skibba15}. The lack of much residual correlation of clustering amplitude with \Ha\ luminosity when stellar mass is fixed is clearly demonstrated in Fig.~\ref{fig:r0_stellar_mass} with similar results for the galaxy bias and central halo mass revealed by the halo modelling in Fig.~\ref{fig:median_bias_median_mcen}. 

We do see evidence that there may be a small residual V-shaped \Ha\ luminosity dependence to the clustering amplitude when stellar mass is fixed, such that the lowest and highest \Ha\ luminosity galaxies in any given stellar mass bin (i.e. highest and lowest sSFR) are slightly more clustered than those in middle. We do not see any evidence of such a trend when we look at how the typical halo mass of central galaxies depends on stellar mass and \Ha\ luminosity in our best fit halo model, with stellar mass being utterly dominant (middle panel of Fig.~\ref{fig:median_bias_median_mcen}). This implies that this residual V-shaped \Ha\ luminosity dependence to the clustering amplitude is being driven by the satellite galaxy population. Such a trend would occur if at fixed stellar mass the satellite fraction is higher at the highest and lowest \Ha\ luminosities and hence sSFRs. Our halo modelling provides some indication of the lowest satellite fractions always falling in the central \Ha\ luminosity bin when stellar mass is fixed (bottom panel of Fig.~\ref{fig:median_bias_median_mcen} and Table~\ref{tab:derived M_*}). This dependence of the satellite fraction on \Ha\ luminosity in the best fitting model arises as a result of a broader scatter in \Ha\ luminosity at fixed halo mass for star-forming satellites ($\sigma_{y,{\rm sat}}$) than for star-forming central galaxies. Physically this makes sense as satellite galaxies may experience an initial enhancement of star formation on infall into a cluster \citep[][e.g.]{Vulcani18} followed by a reduction as their gas supply is removed \citep[see][for a review]{Cortese21}, leading to a wider range in sSFRs than equivalent central galaxies. With these tantalising results, it is important to note that our constraints on the satellite population are weak at best from the halo modelling, but it is noteworthy that they nicely explain the trend seen in basic clustering amplitude measurements in Fig.~\ref{fig:r0_stellar_mass}.     

Throughout this work we have treated our sample of star-forming galaxies with $0.7 < z < 1.5$ as a single population and ignored any potential redshift evolution in their properties over this period\footnote{We do include the redshift evolution in the dark matter halo properties in our halo model, fitting at the mean redshift of each galaxy sample.}. We do know that the \Ha\ luminosity function is evolving \citep{Sobral13} and that there is a small amount of evolution in the star formation main sequence over this redshift range \citep{whitaker,Schreiber15}. In Fig.~\ref{fig:r0_vs_cochrane} we show measurements of the relations between $r_0$ and \Ha\ luminosity in luminosity-threshold (left) and luminosity-bin (middle) samples (Table~\ref{tab:lum_bin_thres}) made exactly as before but split into two redshift ranges, $0.7 < z < 1.1$ and $1.1 < z < 1.5$. While the relations at both redshifts are similar, there is an offset such that the lower redshift galaxies have a higher clustering amplitude at the same \Ha\ luminosity, except for  the highest \Ha\ luminosities. Much of the evolution in the \Ha\ luminosity function over this redshift range can be explained by pure luminosity evolution, i.e. evolution of the characteristic luminosity $L^*$ \citep[Vang et al. in prep]{Sobral13}, and this is also the case for the clustering. In the right panel of Fig.~\ref{fig:r0_vs_cochrane} we show, for the luminosity-bin samples, $r_0$ as a function of \Ha\ luminosity divided by $L^*$ calculated at the mean redshift of each sample, where $L^*(z) = 2.63\times 10^{41}(1+z)^{2.36} {\rm erg\, s^{-1}}$ from the 3D-HST data (Vang et al. in prep). Correcting the \Ha\ luminosity by $L^*$ removes practically all the redshift evolution with the two relations now virtually lying on top of each other.   

Given there is some redshift evolution in the clustering amplitude at fixed \Ha\ luminosity over our redshift range, it is worth considering if that will affect any of our main results. When dividing our samples only by \Ha\ luminosity, we do introduce a redshift trend such that as the \Ha\ luminosity increases so does the mean redshift of the sample, with it going from 0.82 to 1.21 from lowest to highest luminosity respectively. Given the results shown in Fig.~\ref{fig:r0_vs_cochrane}, the redshift evolution will have a tendency to flatten the relationship between $r_0$ and \Ha\ luminosity shown in Fig.~\ref{fig:r0_lum}. There can be a similar tendency in the relation between galaxy bias $b_g$ and \Ha\ luminosity and that between median mass $M_{\rm h,cen}$ of haloes hosting central galaxies and \Ha\ luminosity (Fig.~\ref{fig:derived_lum_bins}). However, we expect the effect to be smaller than that on $r_0$. We can see this by noting that the clustering amplitude is proportional to $b_g^2D(z)^2$ as well as $r_0^\gamma$, where $D(z)$ is the linear growth factor. That is, $b_g\propto r_0^{\gamma/2}/D(z)$. We have $\gamma/2\sim 0.7<1$ and $D(z)$ decreases with increasing redshift -- both factors make $b_g$ less sensitive to the sample redshift than $r_0$, given that we have lower $r_0$ at higher $z$ with fixed \Ha\ luminosity. We note that in our HOD modelling the redshift of each sample is adopted to compute halo properties (e.g. halo bias, halo mass function). Therefore, the derived quantities shown in Fig.~\ref{fig:derived_lum_bins}, including $b_g$ and $M_{\rm h,cen}$, correspond to those at the mean redshift of each sample.

When we bin by stellar mass as well as \Ha\ luminosity, it is the stellar mass binning that dominates in determining the mean redshift of a given sample, with the mean redshift increasing by less than 0.3 between the highest and lowest stellar mass samples at fixed \Ha\ luminosity. Within a given stellar mass bin the mean redshift changes by at most 0.09 with \Ha\ luminosity and so any redshift evolution is negligible. Given that clustering amplitude at fixed stellar mass hardly evolves at all with redshift at these redshifts \citep[e.g.][]{wake,Skibba15}, we can expect the trends shown in Fig.~\ref{fig:r0_stellar_mass} and results of the halo model fitting in Section~\ref{sec:HOD_lum_mass} to be largely unaffected by redshift evolution. To make sure that this is the case we have reproduced Fig.~\ref{fig:r0_stellar_mass} for the two redshift ranges discussed above, confirming that the same trends are observed, but with a larger scatter as would be expected. 

There have been previous analyses of the clustering of \Ha\ emitting galaxies both observationally and in simulations. The most comparable to our study are those made with the  High-Redshift(Z) Emission Line Survey (HiZELS), which is a deep, near-infrared narrow-band \Ha\ survey targeting galaxies in narrow redshift bins at $z=$ 0.8, 1.47, and 2.23 \citep{Geach08,Sobral09, Sobral12, Sobral13}. The \Ha\ emitting galaxies are selected based on the narrow-band and broad-band colour excess, with a restframe \Ha\ equivalent width above 25\AA. Photometric redshifts are used to ensure that the measured emission line is \Ha. Of particular relevance are \citet{cochrane17} and \citet{cochrane18}, which present analyses of the \Ha\ luminosity dependent clustering and \Ha\ luminosity and stellar mass dependent clustering of HiZELS galaxies, respectively. The HiZELS data used in these analyses cover a similar volume as the data we use here. Their two $z\simeq 0.8$ fields ($z=0.845\pm 0.011$ and $z=0.81\pm 0.011$) cover a comoving volume of $4.60\times 10^5 h^{-3}{\rm Mpc}^3$ with their $z\simeq 1.47$ field covering $2.68\times 10^5 h^{-3}{\rm Mpc}^3$, compared to the $3.83\times 10^5 h^{-3}{\rm Mpc}^3$ covered by the 3D-HST survey. 3D-HST extends to lower \Ha\ luminosities than HiZELS, meaning about 50\% more galaxies are available to use in our clustering analysis. 

In Fig.~\ref{fig:r0_vs_cochrane}, we compare the clustering strengths $r_0$ from our work with those in \citet{cochrane17} for the \Ha\ luminosity-threshold (left) and luminosity-bin samples (middle and right). To enable a closer comparison, we have divided our samples into two redshift bins when measuring their clustering, $0.7<z<1.1$ and $1.1<z<1.5$. For the power-law fits, the index is fixed to 1.8 to be consistent with that used in \citet{cochrane17}. \citet{cochrane17} add a dust attenuation correction of $A_{\rm H\alpha} = 1.0$ mag, so we subtract their $\log L_{\rm H\alpha}$ values by 0.4 dex to be in line with the values adopted in our work.

As can be seen in Fig.~\ref{fig:r0_vs_cochrane}, our 3D-HST samples are able to reach a lower \Ha\ luminosity than the HiZELS samples at similar redshifts. Looking at the $r_0$ values we measure for our $0.7<z<1.1$ samples, we see that at lower luminosity they are consistent with those from the HiZELS $z\simeq 0.8$ samples. As we move to higher luminosities, our $r_0$ values lie below those of HiZELS at $z\simeq 0.8$, producing a shallower dependence of $r_0$ on \Ha\ luminosity than observed by HiZELS. This shallower slope is not being caused by redshift evolution within our samples, with the mean redshift varying from 0.93 to 0.95 in the overlapping \Ha\ luminosity range of the two surveys, although there is an overall difference of about 0.1 in redshift with HiZELS. For our higher redshift ($1.1<z<1.5$) samples, the $r_0$ values agree with those from HiZELS $z\simeq 1.47$ samples, but show a much clearer trend as a result of our larger sample size, as well as an extension to lower \Ha\ luminosity. While the results from \citet{cochrane17} show a clear evolution of the clustering strength from $z\simeq 1.47$ to $z\simeq 0.8$ at all \Ha\ luminosities, we see only a small difference for galaxies of the highest luminosity in our sample. This may partially be caused by the larger redshift difference between the two HiZELS samples, $\Delta z \sim 0.65$, compared to our mean redshift difference of $\Delta z \sim 0.4$, but it could also simply be a reflection of the larger cosmic variance uncertainties in the HiZELS data from their smaller number of independent fields.    

\citet{cochrane18} measure the \Ha\ luminosity and stellar mass dependent clustering much as we do in Section~\ref{sec:stellar_mass}, but appear to find quite different results. They conclude that there is evidence for residual clustering variation with stellar mass at fixed \Ha\ luminosity only at the highest stellar masses or \Ha\ luminosities, but there is clear evidence for residual \Ha\ luminosity dependent clustering at fixed stellar mass, such that more luminous galaxies are more strongly clustered. This is quite different to what we concluded from Figs.~\ref{fig:r0_stellar_mass} and \ref{fig:median_bias_median_mcen}. \citet{cochrane18} also find no dependence of $r_0$ on stellar mass for their \Ha\ emitting galaxies of stellar mass below $4\times 10^{10} \Msun$, with an increase above that. That is also different to our results (left panel of Fig.~\ref{fig:r0_stellar_mass}, open circles), where we see the clustering amplitude increase steadily for masses above $9.6 \times 10^9 \Msun$. 

\begin{figure}
  \includegraphics[width =.45\textwidth]{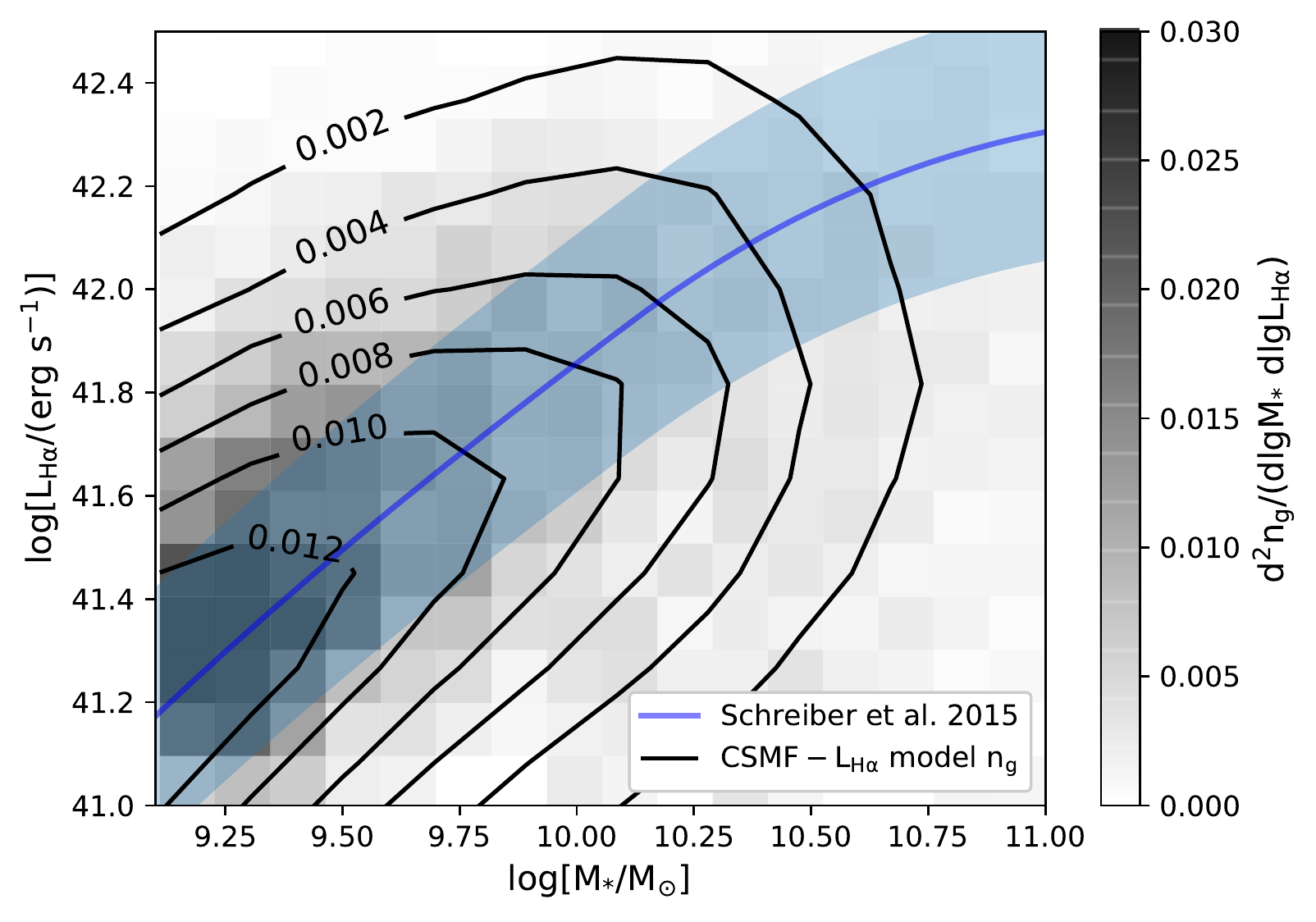}
  \caption{Distribution in the \Ha\ luminosity--stellar mass plane of the star-forming galaxies in our sample (shaded 2D-map), and from our HOD model (black contours). The measured number densities derived from our sample include our $V_{\rm max}$ and completeness corrections but are cut off at our luminosity limit of $10^{41} {\rm erg\, s^{-1}}$. The model distribution is able to extend below the luminosity limit of our data. For comparison, we show the same relation derived from fits to the observed star formation mass sequence of \citet{Schreiber15} (blue curve with shaded area showing the 1$\sigma$ scatter). The \citet{Schreiber15} measurement is in good agreement with our best-fitting model distribution.}
  \label{fig:sfms}
\end{figure}

The most likely cause of these differences are two fold and quite straightforward. Firstly, the HiZELS data has a higher \Ha\ luminosity limit that that of 3D-HST. Our 3D-HST samples go to about 0.3 dex lower in \Ha\ luminosity (e.g. Fig.~\ref{fig:lum_bin_Lha_vs.z}), meaning that we sample below the star formation mass sequence or star formation main sequence (SFMS) over our full stellar mass range. We show in Fig.~\ref{fig:sfms} the distribution of our galaxies in the \Ha\ luminosity--stellar mass plane, along with the same relation (blue curve) derived from fits to the observed SFMS in \citet{Schreiber15}. To convert the SFR in the \citet{Schreiber15} SFMS to \Ha\ luminosity, we use the $L_{\rm H\alpha}$--SFR relation in \citet{Kennicutt12} and then apply the stellar-mass-dependent dust extinction given by \citet{Garn10}. 
We sample below the 1$\sigma$ scatter of the \citet{Schreiber15} SFMS over virtually our full stellar mass range and below 2$\sigma$ for galaxies with $M_* > 4\times10^9 \Msun$. In comparison, \citet{cochrane18} probe above the SFMS at their lowest stellar masses and do not sample galaxies 2$\sigma$ below until masses of $8\times 10^9 \Msun$. By only sampling the highest \Ha\ luminosity (SFR) galaxies at the lowest stellar masses, as \citet{cochrane18} note they likely preferentially selecting satellite galaxies with enhanced star formation (see our Fig.~\ref{fig:median_bias_median_mcen}). They are also more likely to select galaxies that have measured masses lower than their true values as a result of measurement errors \footnote{Given the physical correlation between stellar mass and \Ha\ luminosity, galaxies in a given bin in observed stellar mass with high \Ha\ luminosities are relatively more likely to have true stellar masses higher than that stellar mass bin and have been scattered in by measurement error. Conversely galaxies with low \Ha\ luminosities are relatively more likely to have true stellar masses that are lower than the stellar mass bin they are in.}. Both cases will cause the clustering amplitude to be enhanced for the lowest mass galaxies, reducing any dependence on stellar mass. 

Likewise if we were to increase our \Ha\ luminosity limit by 0.3 dex to match \citet{cochrane18} we would mainly be sampling the higher luminosity rising side of our observed V-shaped trend of $r_0$ with \Ha\ luminosity at fixed stellar mass. That may have led us to conclude that there was evidence of a residual trend of increasing $r_0$ with \Ha\ luminosity independent of stellar mass at the highest \Ha\ luminosities as \citet{cochrane18} have done. 

It is also important to consider the relatively large uncertainties in the $r_0$ measurements from both studies. While some of the trends are different, a close comparison of the individual measurements in overlapping ranges of mass and \Ha\ luminosity show reasonable agreement given those errors. It will take a larger data set, most likely from Euclid or RST, to resolve these differences.

\citet{Coil17} and \citet{Berti19} investigate the sSFR dependent clustering of star-forming and quiescent galaxies at $0.2 < z < 1.2$ from the PRIMUS survey. \citet{Coil17}, looking at the full galaxy population, found some evidence for sSFR dependent clustering in star-forming galaxies with galaxies with lower sSFRs having a somewhat larger clustering amplitude. Our results do not confirm this trend, although we do find our $r_0$ and bias values are entirely consistent with theirs at comparable redshift, stellar masses, and SFRs. The differences may be driven by the choice of \citet{Coil17} to divide their sample by sSFR over a wide range in mass rather than our choice to use fairly narrow bins in each. \citet{Berti19} attempt to study the clustering of central galaxies only using an isolation criteria. Interestingly, these largely central only samples show no significant dependence of the clustering amplitude on sSFR for star-forming galaxies at fixed mass. This implies, much as we have found, that any residual dependence of clustering strength on sSFR is being driven by the satellite galaxy population.

\section{Conclusions}
\label{sec:conclusion}

In this work, using 3D-HST grism spectroscopic survey data, we study the clustering of \Ha-emitting galaxies at $z\sim 1$. The projected 2PCFs of galaxies are measured as a function of \Ha\ luminosity ($L_{\rm H\alpha}$) and stellar mass ($M_*$) and characterised with power-law fits. We develop halo-based models, including the CLF model and the conditional $M_*$--$L_{\rm H\alpha}$ distribution model, to interpret the clustering measurements and infer the relation between \Ha-emitting galaxies and dark matter haloes.

The main results are listed below.
\begin{itemize}
  \item[(1)] We find a clear correlation between \Ha\ luminosity and clustering amplitude, with more luminous galaxies having a higher clustering amplitude. 
  
  \item[(2)] At fixed \Ha\ luminosity galaxies at lower redshift ($0.7 < z < 1.1$) are found to be more clustered than those at higher redshift ($1.1 < z < 1.5$), but the difference in clustering strength between our samples in the two redshift ranges is relatively small. 
  
  \item[(3)] Our measurements of the \Ha\ luminosity dependent clustering broadly agree with those of \cite{cochrane17}, while we extend to lower luminosities.

  \item[(4)] At fixed \Ha\ luminosity, galaxies with higher stellar mass tend to be more strongly clustered. 
  \item[(5)] At fixed stellar mass, clustering strength does not seem to be dependent on \Ha\ luminosity. It implies that the dependence of clustering on \Ha\ luminosity is primarily driven by the relatively tight correlation between \Ha\ luminosity and stellar mass (a.k.a. the star formation main sequence). 
  
  \item[(6)] We use the CLF model to interpret the \Ha\ luminosity dependent clustering. We further develop the conditional $M_*$--$L_{\rm H\alpha}$ distribution model to jointly model the clustering measurements of galaxies in a series of stellar mass and \Ha\ luminosity bins. The models provide good fits to our measurements.
  
  \item[(7)] Based on the modelling results, we find that central galaxies with higher luminosity or higher stellar mass reside in haloes of higher mass. Satellite galaxies tend to reside in more massive haloes than similar centrals. The conditional $M_*$--$L_{\rm H\alpha}$ distribution model shows a tight correlation between central galaxy stellar mass and halo mass and only a weak correlation between central \Ha\ luminosity and halo mass.
  
  \item[(8)] From the conditional $M_*$--$L_{\rm H\alpha}$ distribution model, the median galaxy bias appears to have a strong dependence on stellar mass with at most a weak dependence on \Ha\ luminosity most evident in galaxies of the lowest stellar mass. For central galaxies there is a strong dependence of median mass of hosting haloes on stellar mass and essentially no additional dependence on \Ha\ luminosity. Taken together this implies that there is a small residual dependence of the satellite galaxy halo occupation on \Ha\ luminosity at fixed stellar mass. While the satellite fraction is only loosely constrained with the samples we study, we do see an indication that for galaxies of a given stellar mass the satellite fraction increases for both the high and low \Ha\ luminosity galaxies, which is caused by the environment broadening of the range in SFR hence \Ha\ luminosity of satellite galaxies at fixed stellar mass. 
  
\end{itemize}

Our results help inform future surveys of star-forming or emission line galaxies (ELGs), such as the High Latitude Spectroscopic Survey (HLSS) of RST \citep{RST15} and the Euclid survey \citep{Euclid11}. They will be useful in designing the surveys and testing analysis pipeline, e.g. by providing inputs for mock construction. The modelling formalism can be applied to these surveys, as well as existing and ongoing surveys, such as ELG samples in eBOSS \citep{Dawson16} and in DESI \citep{DESI16}. 
 With large samples from these surveys, the formalism like the conditional $M_*$--$L_{\rm H\alpha}$ distribution will be an efficient model to explore the galaxy-halo relation, and we expect to obtain tight constraints on the relation to learn more about star-forming galaxies.


\section*{Acknowledgements}
We would like to thank the anonymous referee for their helpful and constructive comments. 
CC acknowledges the support by a department Swigart Summer Research Fellowship.
DW acknowledges support from program number HST-AR-13274, provided by NASA through a grant from the Space Telescope Science Institute, which is operated by the Association of Universities for Research in Astronomy, Incorporated, under NASA contract NAS5-26555. 
ZZ is supported by NSF grant AST-2007499.
The support and resources from the Center for High Performance Computing at the University of Utah are gratefully acknowledged.
This work is based on observations taken by the 3D-HST Treasury Program (GO 12177 and 12328) with the NASA/ESA HST, which is operated by the Association of Universities for Research in Astronomy, Inc., under NASA contract NAS5-26555.

\section*{Data Availability}

The data underlying this article are available in The STScI 3D-HST Repository, at https://archive.stsci.edu/prepds/3d-hst/. The measurements and modelling results underlying this article are available either in the article or on reasonable request to the authors.



\bibliographystyle{mnras}
\bibliography{references} 


\appendix
\section{Best-fitting Parameters and Derived Quantities}

In this paper, we model the \Ha\ luminosity-dependent galaxy clustering within the CLF framework and then jointly model the stellar mass and \Ha\ luminosity dependent galaxy clustering within the framework of the conditional stellar mass and \Ha\ luminosity distribution. 

\begin{figure}
  \includegraphics[width=.4\paperwidth]{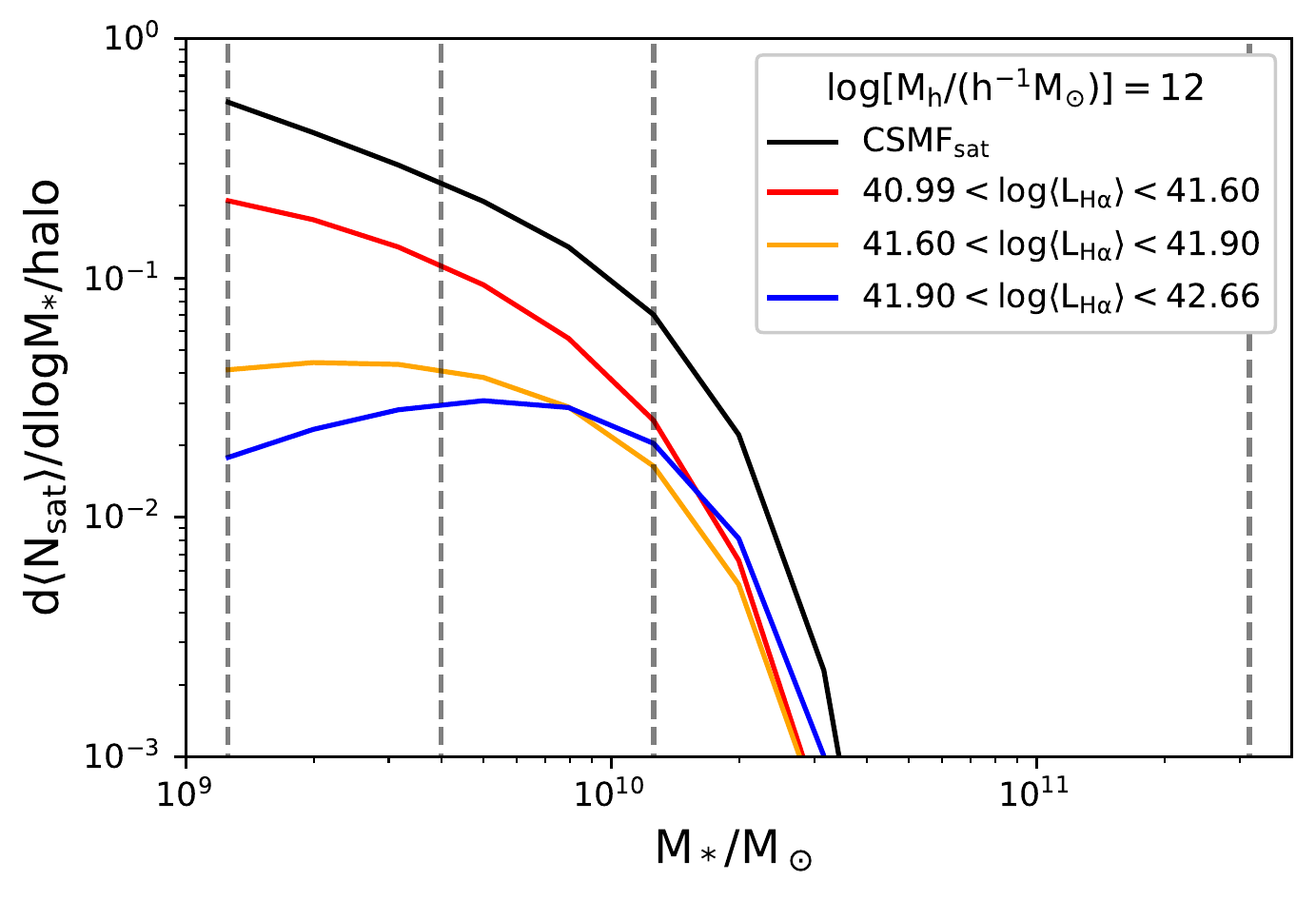}
  \caption{Illustration of the conditional stellar mass and \Ha\ luminosity distribution of satellite galaxies in $10^{12}\hinvMsun$ haloes. The black curve is the conditional stellar mass function (CSMF) of satellites. The red, orange, and blue curves are the CSMF after we apply the \Ha\ luminosity cuts used in this paper. The four vertical lines delineate the three stellar mass bins used in constructing the stellar mass and \Ha\ luminosity bin samples.  
    }
  \label{fig:csmf}
\end{figure}

For the conditional stellar mass and \Ha\ luminosity distribution, at fixed halo mass, that for central galaxies is parameterised as a 2D Gaussian distribution, which is easy to be visualised. That for satellite galaxies is parameterised through the conditional stellar mass function and a relation between \Ha\ luminosity and stellar mass (i.e. following the star formation main sequence). We use Fig.~\ref{fig:csmf} to provide an illustration of the satellite component in $10^{12}\hinvMsun$ haloes. The black curve is the conditional stellar mass function of satellites in these haloes. As we parameterise the \Ha\ luminosity distribution as a function of stellar mass, when applying a cut in \Ha\ luminosity, we obtain the conditional stellar mass function of satellites within the given \Ha\ luminosity bin. The red, orange, and blue curves show the cases for the three \Ha\ luminosity bins used in this paper. The four dashed vertical lines delineate the three stellar mass bins used in constructing the samples. Integrating each of the red, orange, and blue curve over each stellar mass range gives the mean satellite occupation number in haloes of $10^{12}\hinvMsun$ for each stellar mass-\Ha\ luminosity sample. At fixed stellar mass, the  satellite occupation number is not necessarily monotonic with \Ha\ luminosity, and the trend depends on both \Ha\ luminosity cuts and stellar mass.

\begin{figure*}
  \centering
  \includegraphics[width=.8\paperwidth]{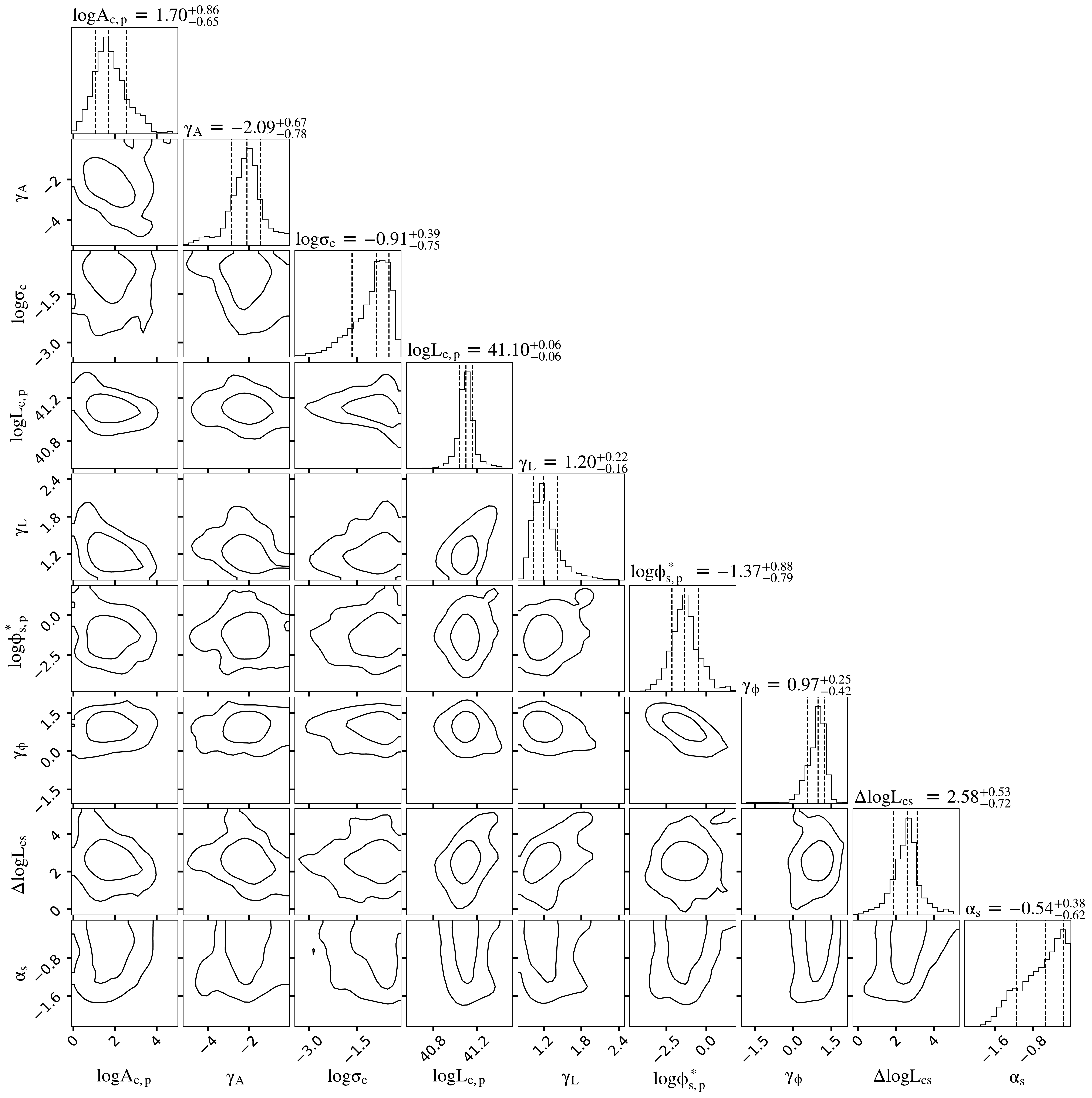}
  \caption{Constraints on the CLF model parameters from modelling $L_{\rm H\alpha}$-bin galaxy samples. Here, $\gamma_A\equiv {\rm d}\log A_{\rm c}/{\rm d}\log \Mh$, $\gamma_L\equiv {\rm d}\log L_{\rm c}/{\rm d}\log \Mh$, and $\gamma_\phi \equiv {\rm d}\log \phi_{\rm s}^*/{\rm d}\log \Mh$. Contours in each panel denote the $1\sigma$ and $2\sigma$ constraints for the pair of parameters. In each histogram panel, the central vertical line marks the median and the other two indicate the central $68.3$ per cent distribution. 
  }
  
  \label{fig:corner_lum_bins}
\end{figure*}

\begin{figure*}
  \includegraphics[width=.8\paperwidth]{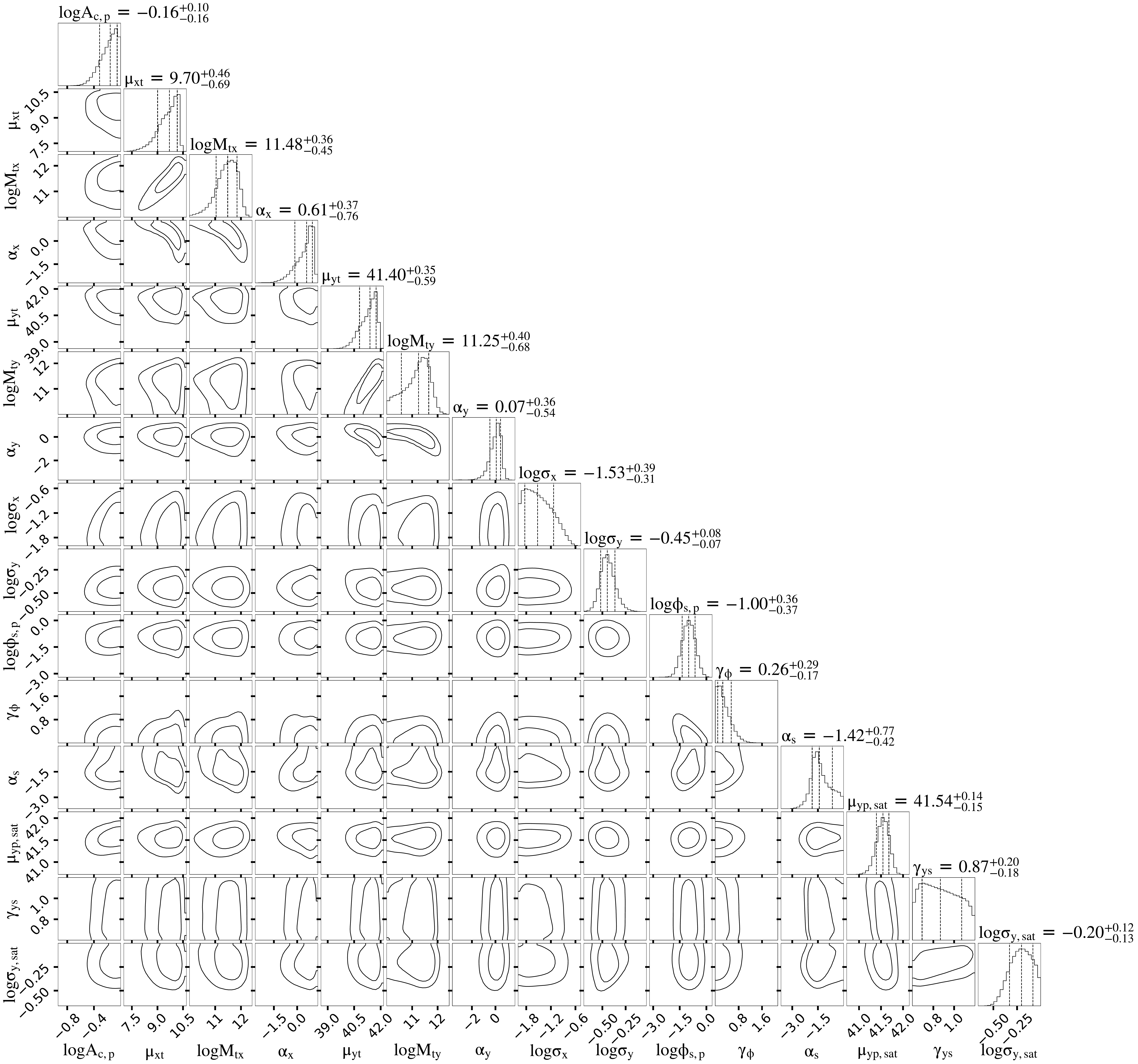}
  \caption{Constraints on the parameters of the conditional stellar mass and \Ha\ luminosity distribution model from jointly modelling stellar mass and \Ha\ luminosity dependence galaxy clustering. Here, $\gamma_\phi \equiv {\rm d}\log \phi/{\rm d}\log \Mh$, and $\gamma_{ys} \equiv {\rm d}\log \mu_{\rm yp,sat}/{\rm d}\log \Mh$. 
  Contours in each panel denote the $1\sigma$ and $2\sigma$ constraints for the pair of parameters. In each histogram panel, the central vertical line marks the median and the other two indicate the central $68.3$ per cent distribution. 
  }
  \label{fig:corner_m*}
\end{figure*}

The parameter constraints for the CLF model and the conditional stellar mass and \Ha\ luminosity distribution model are shown in Fig.~\ref{fig:corner_lum_bins} and Fig.~\ref{fig:corner_m*}, respectively. For the CLF model, the best-fitting parameters with 1$\sigma$ uncertainties and the values of $\chi^2$ are displayed in Table~\ref{tab:CLFpar}. 

Finally, Tables~\ref{tab:derived} and \ref{tab:derived M_*} list the derived parameters from the two models, including the median masses of host haloes for central galaxies and satellite galaxies, the galaxy bias factor, and the satellite fraction. 

\begin{table}
 \caption{Best-fitting CLF parameters for modelling \Ha\ luminosity-bin galaxy samples.}
 \begin{tabular}{||c c c c c||} 
 \hline
 $\log \Acp$ & $\frac{{\rm d}\log \Ac}{{\rm d}\log \Mh}$ & $\log \sigc$ & $\log \Lcp$ & $\frac{{\rm d}\log \Lc}{{\rm d} \log \Mh}$ \\
 \hline 
 $1.70^{+0.86}_{-0.65}$ & $-2.09^{+0.67}_{-0.78}$ & $-0.91^{+0.39}_{-0.75}$ & $41.10^{+0.06}_{-0.06}$ & $1.20^{+0.22}_{-0.16}$ \\
 \hline\hline
$\log \phisp$ & $\frac{{\rm d}\log \phis}{{\rm d}\log \Mh}$ &  $\Delta\log L_{\rm cs}$ & $\alphas$ & $\rm \chi^2/{\rm d.o.f.}$\\
\hline 
$-1.37^{+0.88}_{-0.79}$ & $0.97^{+0.25}_{-0.42}$ & $2.58^{+.53}_{-0.72}$ & $-0.54^{+0.38}_{-0.62}$ & $21.3/57$\\ \hline
\end{tabular}
\label{tab:CLFpar}\\

The CLF parameterisation and parameter definitions can be found in Section~\ref{sec:CLF}.
\end{table}

\begin{table}
 \caption{Derived quantities for \Ha\ luminosity-bin and luminosity-threshold galaxy samples}
 
\begin{tabular}{||c c c c c||}
 \hline
 Sample & $\log M_{\rm h, cen}$ & $\log M_{\rm h, sat}$  & $b_{\rm g}$ & $f_{\rm sat}$\\
 \hline \hline
LB1 & $11.03^{+0.06}_{-0.09}$ & $13.49^{+0.24}_{-0.35}$ & $0.94^{+0.03}_{-0.03}$ & $0.04^{+0.02}_{-0.01}$ \\ \\ 
LB2 & $11.20^{+0.07}_{-0.11}$ & $13.60^{+0.24}_{-0.35}$ & $1.03^{+0.03}_{-0.04}$ & $0.03^{+0.02}_{-0.01}$ \\ \\ 
LB3 & $11.36^{+0.10}_{-0.13}$ & $13.71^{+0.25}_{-0.35}$ & $1.12^{+0.04}_{-0.05}$ & $0.03^{+0.02}_{-0.01}$ \\ \\ 
LB4 & $11.53^{+0.12}_{-0.17}$ & $13.84^{+0.25}_{-0.37}$ & $1.22^{+0.07}_{-0.06}$ & $0.02^{+0.02}_{-0.01}$ \\ \\ 
LB5 & $11.69^{+0.15}_{-0.21}$ & $13.97^{+0.26}_{-0.38}$ & $1.30^{+0.09}_{-0.08}$ & $0.02^{+0.02}_{-0.01}$ \\ \\ 
LB6 & $11.85^{+0.21}_{-0.25}$ & $14.14^{+0.26}_{-0.40}$ & $1.43^{+0.13}_{-0.11}$ & $0.02^{+0.03}_{-0.01}$ \\ \\ 

\hline \hline
LB1Lz & $11.00^{+0.07}_{-0.08}$ & $13.09^{+0.74}_{-1.45}$ & $0.88^{+0.04}_{-0.02}$ & $0.00^{+0.03}_{-0.00}$ \\ \\ 
LB2Lz & $11.16^{+0.10}_{-0.11}$ & $13.12^{+0.70}_{-1.34}$ & $0.95^{+0.03}_{-0.03}$ & $0.00^{+0.02}_{-0.00}$ \\ \\ 
LB3Lz & $11.32^{+0.12}_{-0.15}$ & $13.18^{+0.67}_{-1.26}$ & $1.00^{+0.04}_{-0.04}$ & $0.00^{+0.02}_{-0.00}$ \\ \\ 
LB4Lz & $11.47^{+0.16}_{-0.19}$ & $13.25^{+0.66}_{-1.20}$ & $1.05^{+0.06}_{-0.05}$ & $0.00^{+0.03}_{-0.00}$ \\ \\ 
LB5Lz & $11.61^{+0.20}_{-0.22}$ & $13.32^{+0.66}_{-1.12}$ & $1.10^{+0.08}_{-0.07}$ & $0.00^{+0.04}_{-0.00}$ \\ \\ 
LB6Lz & $11.77^{+0.23}_{-0.25}$ & $13.42^{+0.67}_{-1.03}$ & $1.20^{+0.12}_{-0.10}$ & $0.01^{+0.07}_{-0.01}$ \\ \\ 
LB1Hz & $11.32^{+0.13}_{-0.15}$ & $12.24^{+0.87}_{-0.90}$ & $1.15^{+0.06}_{-0.06}$ & $0.02^{+0.07}_{-0.02}$ \\ \\ 
LB2Hz & $11.42^{+0.16}_{-0.17}$ & $12.31^{+0.84}_{-0.89}$ & $1.23^{+0.08}_{-0.08}$ & $0.02^{+0.07}_{-0.02}$ \\ \\ 
LB3Hz & $11.51^{+0.20}_{-0.19}$ & $12.39^{+0.84}_{-0.88}$ & $1.28^{+0.10}_{-0.10}$ & $0.02^{+0.08}_{-0.02}$ \\ \\ 
LB4Hz & $11.62^{+0.25}_{-0.22}$ & $12.51^{+0.83}_{-0.88}$ & $1.38^{+0.14}_{-0.14}$ & $0.03^{+0.15}_{-0.03}$ \\ \\ 

\hline \hline
LT1 & $11.28^{+0.05}_{-0.05}$ & $13.38^{+0.21}_{-0.27}$ & $1.16^{+0.02}_{-0.03}$ & $0.07^{+0.03}_{-0.02}$ \\ \\ 
LT2 & $11.45^{+0.06}_{-0.07}$ & $13.36^{+0.23}_{-0.27}$ & $1.25^{+0.03}_{-0.04}$ & $0.08^{+0.04}_{-0.02}$ \\ \\ 
LT3 & $11.61^{+0.09}_{-0.09}$ & $13.34^{+0.26}_{-0.27}$ & $1.35^{+0.05}_{-0.04}$ & $0.08^{+0.04}_{-0.03}$ \\ \\ 
LT4 & $11.76^{+0.13}_{-0.12}$ & $13.35^{+0.29}_{-0.27}$ & $1.44^{+0.07}_{-0.06}$ & $0.08^{+0.03}_{-0.03}$ \\ \\ 
LT5 & $11.89^{+0.17}_{-0.15}$ & $13.38^{+0.32}_{-0.27}$ & $1.53^{+0.09}_{-0.07}$ & $0.08^{+0.04}_{-0.03}$ \\ \\ 
LT6 & $12.02^{+0.21}_{-0.16}$ & $13.41^{+0.35}_{-0.27}$ & $1.64^{+0.09}_{-0.08}$ & $0.09^{+0.05}_{-0.04}$ \\ \\ 
\hline 
\end{tabular}
\label{tab:derived}

The derived quantities are the median mass $M_{\rm h,cen}$ of host haloes for central galaxies, the median mass $M_{\rm h,sat}$ of host haloes for satellite galaxies, the galaxy bias factor $\bg$, and the satellite fraction $f_{\rm sat}$. Halo mass is in units of $\hinvMsun$. Priors imposed when modelling stellar-mass-\Ha-luminosity-bin samples lead to higher satellite fractions than those from modelling luminosity-bin samples. The definitions and properties of the samples are found in Table~\ref{tab:lum_bin_thres}.
\end{table}

\begin{table}
 \caption{Derived quantities for stellar-mass-\Ha-luminosity-bin galaxy samples}
 
\begin{tabular}{||c c c c c||}
 \hline
 Sample & $\log M_{\rm h, cen}$ & $\log M_{\rm h, sat}$  & $b_{\rm g}$ & $f_{\rm sat}$\\
 \hline \hline
M1L1 & $11.30^{+0.11}_{-0.14}$ & $12.02^{+0.33}_{-0.38}$ & $1.09^{+0.11}_{-0.11}$ & $0.26^{+0.22}_{-0.18}$ \\ \\ 
M1L2 & $11.32^{+0.11}_{-0.14}$ & $11.99^{+0.28}_{-0.35}$ & $1.17^{+0.10}_{-0.09}$ & $0.17^{+0.19}_{-0.12}$ \\ \\ 
M1L3 & $11.34^{+0.11}_{-0.14}$ & $11.99^{+0.27}_{-0.34}$ & $1.24^{+0.17}_{-0.11}$ & $0.24^{+0.27}_{-0.17}$ \\ \\ 
M2L1 & $11.56^{+0.10}_{-0.13}$ & $12.18^{+0.29}_{-0.26}$ & $1.16^{+0.08}_{-0.06}$ & $0.20^{+0.13}_{-0.10}$ \\ \\ 
M2L2 & $11.59^{+0.10}_{-0.13}$ & $12.17^{+0.26}_{-0.24}$ & $1.23^{+0.06}_{-0.05}$ & $0.13^{+0.09}_{-0.06}$ \\ \\ 
M2L3 & $11.61^{+0.10}_{-0.14}$ & $12.16^{+0.25}_{-0.23}$ & $1.31^{+0.09}_{-0.07}$ & $0.17^{+0.11}_{-0.08}$ \\ \\ 
M3L1 & $12.05^{+0.13}_{-0.14}$ & $12.52^{+0.25}_{-0.27}$ & $1.38^{+0.09}_{-0.09}$ & $0.12^{+0.09}_{-0.06}$ \\ \\ 
M3L2 & $12.06^{+0.10}_{-0.13}$ & $12.53^{+0.23}_{-0.26}$ & $1.49^{+0.07}_{-0.08}$ & $0.08^{+0.06}_{-0.04}$ \\ \\ 
M3L3 & $12.11^{+0.12}_{-0.15}$ & $12.58^{+0.21}_{-0.28}$ & $1.66^{+0.09}_{-0.12}$ & $0.13^{+0.08}_{-0.07}$ \\ \\ 

 \hline
\end{tabular}
\label{tab:derived M_*}

Same as in Table~\ref{tab:derived}, but for stellar-mass-\Ha-luminosity-bin samples. The definitions and properties of the samples are found in Table~\ref{tab:mass_bin_lum_sample}.
\end{table}



\bsp	
\label{lastpage}
\end{document}